\documentclass[usenatbib]{mn2e}
\usepackage{graphicx,times}
\usepackage{amsmath}
\usepackage{amsfonts}
\usepackage{amssymb}
\usepackage{epsfig}
\usepackage[dvips]{color}
\usepackage[usenames,dvipsnames]{xcolor}
\date{\today,~ $ $Revision: 0.9 $ $}
\def\la{\langle}
\def\ra{\rangle}
\def\n{\noindent}
\def\nn{\nonumber}
\def\oh{\hat\Omega}

\def\bk{{\bf k}}

\def\br{{\bf r}}

\def\ad{{_\delta}}

\def\bk{{\bf k}}

\def\bx{{\bf x}}

\def\2p{{(2\pi)^2}}

\def\be{\begin{equation}}
\def\ee{\end{equation}}

\def\beq{\begin{equation}}
\def\eeq{\end{equation}}

\def\ben{\begin{eqnarray}}
\def\een{\end{eqnarray}}

\def\oh{{\hat\Omega}}
\def\nn{{\nonumber}}

\newcommand{\beqa}{\begin{eqnarray}}
\newcommand{\eeqa}{\end{eqnarray}}

\begin{document}
\onecolumn
\title[Non-Gaussianity in Large Scale Structure and Minkowski Functionals]
{Non-Gaussianity in Large Scale Structure and Minkowski Functionals}
\author[Pratten \& Munshi]
{Geraint Pratten$^{1}$, Dipak Munshi$^{1}$\\
$^{1}$School of Physics and Astronomy, Cardiff University, Queen's
Buildings, 5 The Parade, Cardiff, CF24 3AA, UK}
\maketitle
\begin{abstract} 
Minkowski Functionals (MFs) are topological statistics that have become one of
many standard tools used for investigating the statistical properties of cosmological
random fields. They have found regular
 use in studies of departures from
Gaussianity in a number of important cosmological scenarios. Important examples include the
Cosmic Microwave Background (CMB), weak lensing studies, 21cm surveys and large
scale structure (LSS). To lowest order the MFs depend on three {\it
generalised} skewness parameters that can be shown to
probe the bispectrum with differing weights. Recent studies have advocated the
use of a
power spectrum associated with the bispectrum, called the {\it skew-spectrum},
that has more power to
distinguish between various contributions to the bispectrum than the
conventional formalism adopted when using the Minkowski Functionals. 
In this article we review
the motivations for studying non-Gaussianity
and emphasize the importance of the momentum dependence of higher order
correlators in investigating both inflationary and early Universe models as well as analytical models for 
gravitational instability. We then introduce the skew-spectra, applied to galaxy surveys, as a tool for
investigating various models for primordial and gravitationally induced non-Gaussianities.
We present analytical expressions for the skew-spectra for the density field and
divergence of the velocity field
in 3D and for projected surveys as a function of redshift and a smoothing angular
scale. A Gaussian window function is assumed throughout this paper.
Analytical results are derived for the case of gravitationally induced non-Gaussianity. 
These results can be generalised to incorporate redshift space effects. This will be
useful in
probing primordial and gravitationally induced non-Gaussianity from ongoing and future galaxy surveys.
\end{abstract}
\section{Introduction}
\label{sec:int}
The canonical model for inflation is a single free scalar field obeying the Hamilton-Jacobi slow-roll constraints in a semiclassical background governed 
by Einstein gravity with initial conditions set by assuming that the vacuum state asymptotically approaches the Bunch-Davies vacuum.
The appeal of invoking such an inflationary model is that it provides a rather natural mechanism for 
solving the causality and flatness problems whilst allowing for the generation of primordial density perturbations through quantum mechanical fluctuations
\citep{Mukhanov81,Starobinsky82,Hawking82,Guth82}. 
The scalar field model draws motivation from concepts in particle physics and in many ways represents a vanilla model for inflation in the sense that we have 
made rather minimal assumptions about the presence of physics beyond the standard model in the early Universe. It should be emphasized, however, that whilst 
the scalar field models draw on ideas from particle physics there is, as of yet, no known candidate particle that could source the hypothetical quantum field 
driving inflation (see, for example, \cite{Lyth99}, \cite{Linde05} and \cite{Mazumdar11} for reviews of particle physics models for inflation). 
The energy scales typically associated with inflationary cosmology are currently far beyond that accessible to terrestrial experiments 
and we must necessarily extrapolate current theories to these energy scales prompting questions over the validity of the effective field theories in 
describing inflationary models and possible quantum gravity (i.e. UV) inspired corrections \citep{Easther01, Martin01, Danielsson02, Kachru03, Baumann09}. 

The CMB probes the early Universe in a 
linear regime acting, currently, as one of the most direct probes of the dynamics of the early Universe. One of the largest problems with current and future 
CMB surveys (e.g. polarization) is the relatively small signal-to-noise ratios. Planck is expected to exhaust the information content of the temperature 
anisotropies but will not provide strong statistical constraints on the polarization with a number of dedicated polarization missions in the development stage 
(e.g. CMBPol, COrE). Given that we only have access to one realization of the Universe, it is important to use as many cosmological data sets as possible to 
analyse the statistical properties of primordial density perturbations 
(e.g. \cite{Munshi11,Mead11,Nicholson09}). 
\n
A number of recent papers have suggested that the statistics of dark halos may provide as stringent constraints on primordial non-Gaussianity as 
CMB observations (e.g. \cite{Carbone08, Dalal08, Matarrese08}). When observing large scale structure we encouter more severe 
non-linearity and we find that there are a number mechanisms for generating significant non-Gaussianity. The first 
and most prominent is the non-linear evolution of structure through gravitational instability. A second source of non-linearity is the bias relation 
between the galaxy and matter distributions that arises when considering large scale structure surveys. The final, and in our case most interesting, 
source of non-Gaussianity is primordial in origin. In this paper we consider the role of large scale structure (LSS) surveys in 
providing a probe of early Universe cosmology and further develop the skew-spectra as a tool to distinguish between various contributions to the observed 
bispectrum of LSS.

The MFs have been extensively developed as a statistical tool in a cosmological setting for both 2-dimensional (projected) and 3-dimensional (redshift) surveys. The MFs have analytically known results for a Gaussian random field making them suitable for studies of non-Gaussianity. Examples of such studies include CMB data (\cite{Natoli10,HikageM08,Novikov00,Schmalzing98}), weak lensing (\cite{Matsubara01,Sato01,Taruya02,MunshivW11}), large scale structure (\cite{Gott86,Coles88,Gott89,Melott89,Moore92,Gott92,Canavezes98,Schmalzing00,Kerscher01,Park05,Hikage08,Hikage06,Hikage02}), 21cm (\cite{Gleser06}), Sunyaev-Zel'dovich (SZ) maps (\citep{MunshiSZ11}) and N-body simulations (\cite{Schmalzing00,Kerscher01}). Note that this is an incomplete list of references and we have selected a sample of representative papers from the literature. The MFs are spatially defined topological statistics and, by definition, contain statistical information of all orders. This makes them complementary to the polyspectra methods that are defined in Fourier space. It is also possible that the two approaches will be sensitive to different aspects of non-Gaussianity and systematic effects although in the weakly non-Gaussian limit it has been shown that the MFs reduce to a weighted probe of the bispectrum (\cite{Hikage06}).

The next decade should see the next generation of large scale structure surveys beginning to produce cosmological data sets in unprecedented detail with the most notable of these surveys including: WiggleZ \footnote{http://wigglez.swin.edu.au/site/} \citep{Blake11}, Euclid \footnote{http://euclid.gsfc.nasa.gov/} \citep{Laureijs09}, WFIRST \footnote{http://wfirst.gsfc.nasa.gov/} \citep{Green11}, Boss \footnote{http://cosmology.lbl.gov/BOSS/}, BigBoss \footnote{http://bigboss.lbl.gov/} \citep{Schlegel10} and Subaru PFS \footnote{http://sumire.ipmu.jp/en/}. These future surveys can be combined with the latest CMB measurements to effectively trace the growth of structure formation from the surface of last scattering ($z \approx 1100$) through to the local Universe ($z \sim \mathcal{O}(1)$). The availability of such data sets promises to provide tight constraints on cosmological parameters (\cite{Wang99,Tegmark04a}), angular diameter distance derived from BAO measurements (\cite{Percival07}), Hubble expansion rate, growth rate of structure formation (\cite{Percival09}), Gaussianity of initial conditions and the nature of late time cosmic acceleration (\cite{Albrecht06}). With more comprehensive data sets, a deeper understanding of systematics (e.g. \cite{Hikage11}) and improved understanding of both primordial and late time contributions to non-linearity of the growth of structure (e.g. \cite{Bernardeau02,Chen10}) it becoming possible to do comprehensive studies utilising higher order statistics. In particular we are slowly reaching a point where it will be possible to break degeneracies between various contributions to observed non-linearity (primordial, gravitational instability, bias, systematics, etc) and constrain theories for the early and late Universe.

This paper is organised as follows. In \textsection\ref{sec:Mink} we briefly review the concept of Minkowski Functionals in 3D and 2D as 
a morphological descriptor. In \textsection\ref{sec:skew_spec} we introduce the skew-spectra. The next section \textsection\ref{sec:gi_pm}
is devoted to gravity induced non-Gaussianity as well as primordial non-Gaussianity. The section \textsection\ref{sec:3Dden} introduced
the skew-spectra for a 3D field in general and galaxy distribution in particular. Next, in section  \textsection\ref{sec:3Dvel} we 
develop the theory for skew-spectra for the divergence of 3D velocity field. The projected skew-spectra are considered in  \textsection\ref{sec:2D}.
Finally  \textsection\ref{sec:conclu} is devoted to discussion of our results and conclusion.

The particular cosmology that we will adopt for numerical
study is specified by the following parameter values (to be introduced later):
$\Omega_\Lambda = 0.741, h=0.72, \Omega_b = 0.044, \Omega_{\rm CDM} = 0.215,
\Omega_{\rm M} = \Omega_b+\Omega_{\rm CDM}, n_s = 0.964, w_0 = -1, w_a = 0,
\sigma_8 = 0.803, \Omega_\nu = 0$.

\section{Minkowski Functionals: Integral Geometric Tools as a Probe of Non-Gaussianity}
\label{sec:Mink}
In this section we provide a brief introduction to integral geometry and define the Minkowski Functionals for a d-dimensional space.
Integral geometry provides a natural framework for the set of morphological descriptors of a given random field. These descriptors are intrinsically defined
in the spatial domain and naturally take into account all n-point correlators up to arbitrary order. This provides a nice complementarity between the Fourier based 
methods and the spatially defined topological statistics. Hadwiger's characterisation theorem 
shows that a linear combination of a set of d+1 functionals will provide a complete morphological description of our random field in a d-dimensional space \citep{Hadwiger59}. 
These functionals are more commonly referred to as the Minkowski functionals (MFs) and are defined, adopting an integral geometric approach, 
over a convex ring (i.e. any finite union
of compact, convex sets $\mathcal{K}_d$ in a d-dimensional Euclidean space $\mathbb{E}^d$). The importance of Minkowski Functionals is that they will be the unique morphological descriptors that obey motion-invariance, convex-continuity and additivity. 
These properties are desirable in a morphological descriptor due to the interpretation of these properties when working on, for example, pixelised data sets 
(see e.g. \cite{Winitzki98}).
The Minkowski functionals are commonly calculated using volume weighted curvature integrals and the appeal of this formalism is that there exist 
exact analytical results for an underlying Gaussian random field \citep{Adler81,Tomita86,Gott90}. 
More recently the analytical results have been extended to weakly non-Gaussian random fields through a perturbative approach based on an 
Edgeworth expansion (e.g. \cite{Matsubara94,Matsubara95,MatsubaraY96,Matsubara02,Hikage06}) as a function of a set of skewness parameters. These results allow us to 
use the MFs as a test for non-Gaussianity in the perturbatively weak regime. This is convenient as observations from the CMB and LSS constrain 
non-Gaussianity to be small and, importantly, the levels of non-Gaussianity generated by models for inflation, 
reheating, preheating, secondary effects and gravitational instability are typically well within the domain of validity for the perturbative results \citep{Hikage06}. \\

\n
The MFs can be defined for both 2D (projected) and 3D (redshift) data sets. 
The MFs have been applied to CMB data sets \citep{Coles88,Park88,Colley96,Novikov00,Komatsu03,Eriksen04,HikageM08,Natoli10}, 
large scale structure 
\citep{Gott86,Gott89,Melott89,Coles91,Moore92,Gott92,Mecke94,Rhoads94,Schmalzing97,Canavezes98,Sahni98,Schmalzing99,
Schmalzing00,Park05,Hikage02,Hikage03,Hikage06,Hikage08} and weak lensing surveys 
\citep{MunshivW11}. The MFs, being intrinsically defined in the spatial domain, provide a probe of all orders of correlation functions in contrast to 
the more conventional 
polyspectra or Fourier-space methods (e.g. power spectrum, bispectrum and trispectrum). In the weakly-non-Gaussian limit it was shown \citep{Hikage06} that 
the MFs reduce 
to a weighted probe of the bispectrum given in terms of a set of skewness parameters. This makes the MFs complementary as it offers an alternative 
probe of the data in the presence of contaminants such as survey masks, inhomogeneous noise, foregrounds, etc \citep{Hikage06}.
Whilst the CMB offers, perhaps, the cleanest probe of non-Gaussianity 
large scale structure surveys offer a complementary data set that can be used to investigate the nature of primordial non-Gaussianity and the nature 
of gravitational instability associated with the growth of structure. \\
%

\n
In this paper we will consider the Minkowski Functionals for 3D random fields as well as 2D projected surveys in the context of large scale structure. A 3-dimensional
cosmological density field can be treated as a smoothed scalar field $\Psi(x)$ with zero mean $\la \Psi(x) \ra = 0$. The excursion set, $Q$,
can be calculated over a given threshold, $\nu$, by taking all the points of the random field above the threshold: $Q = \lbrace {\bf{x}} | \Psi({\bf{x}}) > \nu
\sigma \rbrace$, where $\sigma = \la \Psi^2 \ra^{1/2}$ is the standard deviation of the field. For a three-dimensional space the Minkowski Functionals can be 
defined using curvature $(\kappa_1,\kappa_2)$ weighted integrals:
\begin{eqnarray}
\label{eqn:CurvatureMFs}
V_{0} (\nu) = \frac{1}{V} \int_{V} d^3 x \, \Theta (\nu \sigma - \Psi({\bf{x}}));\;
 V_{1} (\nu) = \frac{1}{6 V} \int_{\partial Q} ds; \;
  V_{2} (\nu) = \frac{1}{6 \pi V} \int_{\partial Q} ds \left[ \kappa_1 ({\bf{x}}) + \kappa_2 ({\bf{x}}) \right];\;  
 V_{3} (\nu) = \frac{1}{4 \pi V} \int_{\partial Q} ds \, \kappa_1 ({\bf{x}}) \kappa_2 ({\bf{x}}).
\end{eqnarray}
\n
The formula for MFs in both a Gaussian \citep{Tomita86} and
a weakly non-Gaussian field \citep{Matsubara02} are analytically known allowing us to compare observational and numerical results to expected values. The generalisation to a 2D field is trivial and follows the same approach but of reduced dimensionality. \\

\n
Following the notation of
\cite{Hikage06}, the formula for a MF in a weakly non-Gaussian field is given as a function of the threshold, $\nu$, with an amplitude set by $A_k$: $V_k (\nu) = A_k v_k (\nu)$
The amplitude of the Minkowski Functionals are weighted by the spatial dimension, the variance of the random field, $\sigma_0$, and the variance
of the derivative field, $\sigma_1$:
 $A_k = ({1}/{(2 \pi)^{(k+1)/2}})({\omega_2}/{\omega_{2-k} \omega_k})\left({ \sigma_1}/{\sqrt{2} \sigma_0} \right)^{k}$
with $\omega_k = \pi^{k/2} / \Gamma (k/2+1)$ being the volume of a k-dimensional unit ball.
The normalised Minkowski Functionals, $v_k (\nu)$, can be perturbatively expanded with respect to the variance, $\sigma$, resulting in a Gaussian term and 
the corresponding non-Gaussian contributions consisting of 
higher order corrections weighted by the variance: $v_k(\nu) = v^{(G)}_k (\nu) + v^{(NG)}_k (\nu)$.
The Gaussian term is given by $v^{(G)}_k = e^{-\nu^2 / 2} H_{k-1} (\nu)$.
Working to order $\sigma_0$ the non-Gaussian corrections are given by:
\begin{equation}
  v^{(NG)}_k = e^{-\nu^2 / 2} \left\lbrace \left[ \frac{1}{6} S_{(0)} H_{k+2} (\nu) + \frac{k}{3} S_{(1)} H_k (\nu) + 
\frac{k (k-1)}{6} S_{(2)} H_{k-2} (\nu) \right] \sigma_0 + \mathcal{O} (\sigma^2_0) \right\rbrace ,
\end{equation}
\n where $H_k (\nu)$ are the k-th Hermite polynomials and $S^{(k)}$ are a set of skewness parameters. The skewness parameters are given by
\citep{Matsubara02}:
\begin{equation}
  S_{0} = {{ \la \Psi^3 \ra }\over{\sigma_0^4}}; \qquad S_{1} = {{ \la \Psi^2 \nabla^2 \Psi \ra } \over {\sigma_0^2 \sigma^2_1}};\qquad
S_{2} = {{2 \la |\nabla \Psi |^2\nabla^2 \Psi \ra} \over {\sigma^4_1}};\quad \sigma^2_j = \int \frac{k^2 dk}{2 \pi^2} k^{2 j} P_{\Psi}(k) .
\end{equation}
\n The Fourier coefficients are defined in terms of the density contrast of galaxies $\delta ({\bf{x}},z)$:
$\delta ({\bf{x}},z) = 1/{(2 \pi)^3}\int{ d^3 {\bf{k}}}\tilde{\delta} ({\bf{k}}) e^{i {\bf{k}} \cdot {\bf{x}}}$.
Next we show how to generalize the skewness parameters to the corresponding skew-spectra. The skew-spectra
carry more information about the relevant bispectra from which they are constructed. 
\section{Generalised Skew-Spectra}
\label{sec:skew_spec}
Previous work has resulted in the development of optimal 3-point estimators for simple models of the bispectrum, most notably for the phenomenological
local configuration parameterised by $f^{\rm{loc}}_{\rm{NL}}$. Most of these optimal approaches compresses the data into a single estimate based on 
$f_{\rm{NL}}$ and consequentially
we lose sensitivity in distinguishing between various contributions to the observed non-Gaussianity. 

The skew-spectra are a proposed method utilising cubic statistics constructed from the cross-correlation of two differing fields. These fields are 
constructed as a product of the maps and their derivatives. The three different skewness parameters that we generate will reduce to a weighted probe 
of the bispectrum. Following the method developed in \cite{Munshi09} we can define a power spectrum associated with each of these skewness parameters 
and hence we can associate a power spectrum to each of the MFs. The power spectrum associated with a given MF will, by construction, have the same correspondence 
with the various skew-spectra $S^{(j)}_l$ as the MFs have with the one-point cumulants $S^{(j)}$ \citep{Munshi10}.  

The advantage to the associated power spectra is that we retain more of an ability to distinguish between various models for non-Gaussianity rather 
than collapsing all information into a single estimator (e.g. estimators for $f_{\rm{NL}}$). This is due to the generic momenta dependence of higher order correlators such as the bispectrum.
A previous paper \citep{Munshi10} derived analytic results for topological statistics based on the use of skew-spectra 
\citep{Munshi09} allowing us to relate the analytic skewness parameters to the topological properties of large scale structure.
Each of the generalised skewness parameters or generalised cumulant correlators
can be constructed from triplets of field variables that relate to either the original density contrast or to variables 
constructed from derivatives of the fields:
\begin{equation}
\label{eqn:SkewnessParameters}
{S}_{0} (\bx_1,\bx_2) = \frac{\la \delta^2(\bx_1)\delta(\bx_2) \ra_c}{\sigma^4_0} ;
\quad {S}_{1} (\bx_1,\bx_2) = \frac{\la \delta^2(\bx_1) \nabla^2 \delta(\bx_2) \ra_c}{\sigma^2_0 \sigma^2_1};
\quad {S}_{2} (\bx_1,\bx_2) = \frac{\la \nabla\delta(\bx_1)\cdot\nabla\delta(\bx_1) \nabla^2 \delta(\bx_2) \ra_c}{\sigma^4_1}.
\end{equation}
\n
The Bispectrum $B(k_1,k_2,k_3)$ is defined in the Fourier domain as a
three-point correlation function of the
Fourier coefficients:
\begin{equation}
\la \delta(z)(\bk_1) \delta(z)(\bk_2) \delta(z)(\bk_3) \ra_c = (2 \pi)^3 \delta_D (\bk_1 + \bk_2 + \bk_3) B_g ( k_1,k_2,k_3,z) .
\end{equation}
\n The skewness parameters are calculated by integrating over the bispectrum using the appropriate weights:
\begin{align}
S_{0}(k_2,z) &= {1\over \sigma_0^4}{1\over 4 \pi^2}\int_0^{\infty} \frac{k_1^2 dk_1}{4 \pi^2} \int_{-1}^{1} d\mu \,
B_{\delta}(k_1,k_2,|\bk_1+\bk_2|,z)W(k_1R)W(|\bk_1 + \bk_2|R) \label{eq:S0} , \\
S_{1}(k_2,z) &= {3\over 4\sigma_0^2\sigma_1^2}{1 \over 4\pi^2}\int_0^{\infty} \frac{k_1^2 dk_1}{4 \pi^2}
\int_{-1}^{1} d\mu \, | {\bf{k}}_1 + {\bf{k}}_2 |^2  B_{\delta}(k_1,k_2,|\bk_1+\bk_2|,z)W(k_1R)W(|\bk_1+\bk_2|R)
\label{eq:S1} , \\
S_{2}(k_2,z) &= {9\over 4\sigma_1^4}{1\over 4 \pi^2}\int_0^{\infty} \frac{k_1^2 dk_1}{4 \pi^2} \int_{-1}^1 d\mu
\, ({\bf{k}}_1 \cdot {\bf{k}}_2 ) | {\bf{k}}_1 + {\bf{k}}_2 |^2 B_{\delta}(k_1,k_2,|\bk_1+\bk_2|,z)W(k_1R)W(|\bk_1+\bk_2|R) \label{eq:S2} , \\
S_{i} &= \int_0^{\infty} k^2 dk \; {\cal S}^{i}(k)W(kR); \quad i = \in \lbrace 0,1,2 \rbrace .
\end{align}
\n
with $W(kR)$ an arbitrary window function of smoothing radius $R$. Although the window may be generic the two most commonly adopted forms are the top-hat and Gaussian window functions. We adopt the following notation: $|{\bf{k}}_1 + {\bf{k}}_2| = (k^2_1 + k^2_2 + 2 k_1 k_2 \mu)^{1/2}$ 
and the angular terms are given by $\mu = ( {\bf{k}}_1 \cdot {\bf{k}}_2 ) / ( k_1 k_2 )$. 
In our convention the one-point skewness parameters can be recovered by integrating over the second momenta $k_2$. For example:
\begin{equation}
 S_{(i)} (z) = \int_0^{\infty} k^2_2 dk_2 \; S_{(i)}(k,z).
\end{equation}

In the approach presented above we are able to define a power spectrum associated to the MFs by considering the leading order corrections, $\mathcal{O} (\sigma_0)$, to the Gaussian MFs generalised to our skewness parameters:

\begin{equation}
 v^{(NG)}_m (\nu,k,z) \propto \left\lbrace \frac{1}{6} S_0 (k,z) H_{m + 2} (\nu) + \frac{m}{3} S_1 (k,z) H_m (\nu) + \frac{m (m-1)}{6} H_{m-2} (\nu) S_2 (k,z) \right\rbrace.
\end{equation}

Essentially we have associated to each Minkowski Functional a power spectrum defined in terms of the three skew-spectra. The advantage to this approach is that we can study the contributions to the MFs as a function of the Fourier mode $k$. This is useful as a generic model for non-Gaussianity carries momenta dependence giving us greater distinguishing power than the conventional approach which compresses information from all modes into a single statistic. This compression represents a loss of distinguishing power. The skew-spectra presented above are relatively independent of our choice of background cosmology but will be dependent on the model for non-Gaussianity due to the aforementioned momenta dependence. 

Next we show how to generalize the skewness parameters to corresponding skew-spectra.
%
\section{Non-Gaussianity: Primordial and Gravity Induced}
\label{sec:gi_pm}
%
In the weakly non-linear regime ($\delta \le 1$) the description of
gravitational clustering can be 
described by Eulerian Perturbation Theory (see e.g. \cite{Bernardeau02}). 
As the density contrast at a given scale becomes highly non-linear ($\delta \ge 1$) the perturbative treatment breaks down and we observe an increase 
in the growth of clustering. Perturbative studies of gravitational clustering have previously garnered a lot of attention. Starting with \cite{Peebles80} 
there have been a significant number of attempts to reproduce the observed clustering of a self-gravitating fluid in a cosmological setting, most of which 
adopt a brute force approach using N-body simulations \citep{Bernardeau02}. 
In the perturbative approach, solutions are generated by performing a series expansion with higher order corrections being 
introduced to the Fourier expansion of the linear density contrast under the assumption that the density contrast is less than unity for 
the series to be convergent:
\begin{equation}
\delta({\bf k}) = \delta^{(1)}({\bf k}) + \delta^{(2)}({\bf k}) +
\delta^{(3)}({\bf k}) + \dots; \quad
\delta^{(2)}(k) = \int { d^3k_1 \over 2\pi} \int { d^3 k_2 \over 2\pi}
\delta_D({\bf k_1 + k_2 -k }) F_2(k_1,k_2) \delta^{(1)}({\bf k}_1) 
\delta^{(1)}({\bf k}_2)  .
\end{equation}
\n
The linearized solution for the density field is $\delta^{(1)}({\bf k})$ with the
higher order terms describing corrections to the linear term.
Using a fluid approach, known to be valid at large scales, before shell crossing,
one can write the second order corrections
to the linearized density field by introducing a coupling kernel, $F_2({\bf k_1},{\bf k_2})$. 
Newtonian gravity coupled with the Euler and Continuity equations can be used to solve a
system of non-linear coupled integral-differential equations, in order to generate the kernels $F_2(k_1,k_2)$ and
$F_3(k_1,k_2,k_3)$, by solving perturbatively order by order. The expression for the matter bispectrum can be
written in terms of an effective fitting formula that allows us to interpolate between the quasilinear and highly
nonlinear regimes:
\begin{align}
\label{eqn:MatterBispectrum}
B\ad(\bk_1,\bk_2,\bk_3) &= 2 F_2({\bf k_1}, {\bf k_2}) P\ad(\bk_1)P\ad(\bk_2)
+ {\rm cyc.perm.}; \\
F_2({\bf k_1}, {\bf k_2}) &= {5 \over 7}a(n_{e},k_1)a(n_{e},k_2)+ \left (
 { {\bf k}_1 \cdot {\bf k}_2 \over 2 k_2^2}   +{ {\bf k}_1 \cdot {\bf k}_2 \over
2 k_1^2} \right ) b(n_{e},k_1)b(n_{e},k_2)
+ {2 \over 7} \left ( { {\bf k}_1 \cdot {\bf k}_2 \over k_1 k_2} \right )^2
c(n_{e},k_1)c(n_{e},k_2) .
\end{align}
\n
The coefficients $a({n_{e}},k),b({n_{e}},k)$ and $c({n_{e}},k)$ are defined as
follows: 
\begin{equation}
a(n_{e},k) = { 1 + \sigma_8^{-0.2}(z) \sqrt { (q/4)^{n_{e}+3.5}} \over 1 + 
(q/4)^{n_{e}+3.5}}; \quad
b(n_{e},k) = {1 + 0.4(n_e+3)q^{n_{e}+3} \over 1 + q^{n_{eff}+3} }; \quad
c(n_{e},k) = { (2q)^{n_{e}+3}\over 1 + (2q)^{n_{e}+3.5} } 
\left \{ { 1 + \left ({4.5 \over 1.5 + (n_{e}+3)^4} \right )} \right \} .
\end{equation}
Here $n_{e}$ is the effective spectral slope associated with the linear power
spectra
$n_{e} = d \ln P_\ad / d\ln k $, q is the ratio of a given length scale to the
non-linear
length scale $q=k/k_{nl}$, where ${k^3/2\pi^2}D^2(z)P\ad(k_{nl}) = 1$ and 
$Q_3(n_{e})  = {(4 - 2^{n_{e}})/ (1 + 2^{n_{e}})} $. Similarly
$\sigma_8(z)=D(z)\sigma_8$.
At scales where $q \ll 1$, and the relevant length scales are well 
within the
quasilinear regime, then $a = b = c = 1$ and we recover the tree-level perturbative results.
In the regime where $q \gg 1$, and the length scales under consideration are well
within the 
nonlinear scales, we recover $a= \sigma_r^{-0.2}(z) \sqrt {0.7 Q_3(n_{e})}$ with
$b=c=0.$ In this limit the bispectrum becomes 
independent of configuration and we recover a hierarchical form for the bispectrum. 
The possibility, however, of weak violations of the hierarchical ansatz in the
highly nonlinear
regime is still not clear and can only be determined by higher resolution N-body
simulations when they become available. 
Similar fitting functions for a dark energy dominated Universe 
calibrated against simulations are also available and, at least in the
quasilinear regime,
most of the differences arise due to the linear growth factor \citep{Ma99}.

The analytical modeling of the matter bispectrum presented here is equivalent to
the
the so called halo model predictions (see, for example, \cite{Munshi10} or \cite{MunshivW11} for discussions of the 
perturbative treatment in the context of MFs). 
\subsection{Primordial Non-Gaussianity: Bispectrum}
Much of the interest in primordial non-Gaussianity has
focused on a
phenomenological `{\em local} $f_{NL}$' parametrization in terms of the
perturbative non-linear
coupling in the primordial curvature perturbation \citep{Salopek90,Gangui94,Verde00,Komatsu01}: 
\begin{equation}
\Phi(x) = \Phi_L(x) + f_{NL} ( \Phi^2_L(x) - \langle \Phi^2_L(x) \rangle ) + g_{\rm{NL}} \Phi^3_L (x) + h_{\rm{NL}} \left( \Phi^4_L (x) - 3 
\langle \Phi^2_L (x) \rangle \right) + \dots,
\label{eqn:QuadraticNonGaussian}
\end{equation}
\n
where $\Phi_L(x)$ denotes the linear Gaussian term of the Bardeen curvature and
the amplitudes of the various non-Gaussian contributions are parameterised by the set of variables 
$\lbrace f_{\rm{NL}}, g_{\rm{NL}} , h_{\rm{NL}} , \dots \rbrace$.
In this parameterisation, the leading order non-Gaussian contributions are described by the bispectrum or, in configuration space, the three-point correlator. 
A large number of studies involving primordial non-Gaussianity are based around the bispectrum as it 
will contain complete information (in a statistical sense) regarding $f_{NL}$ \citep{Babich05}. The bispectrum has
been extensively studied \citep{Komatsu05,Creminelli03,Creminelli06,Medeiros06,Cabella06,Smith09}, with
most of these measurements providing convolved estimates of the bispectrum. 
In Fourier space, the local-type bispectrum corresponding to \ref{eqn:QuadraticNonGaussian} has the following form:
\begin{equation}
 B^{loc}_{\delta} (k_1,k_2,k_{12}) = 2 f^{loc}_{\rm{NL}} \left[ P_{\Phi}(k_1) P_{\Phi}(k_2) + \rm{cyc}. \right] .
\label{eq:bi_loc}
\end{equation}
Standard inflationary models predict a primordial power spectrum that obeys a power law of the form: $P^{\Phi} (k) \propto k^{n-4}$, where $n$ is the spectral 
index. We can explicitly relate the linear density contrast to Bardeen's curvature perturbations on the matter dominated era through Poisson's equation 
\citep{Hikage06}:
\begin{equation}
 k^2 \tilde{\Phi}_k T(k) = 4 \pi G_N \rho_m (z) \frac{\delta ({\bf{k}},z)}{(1+z)^2} = \frac{3}{2} \Omega_M H^2_0 \delta ({\bf{k}},z) (1+z) .
\end{equation}
In the above, T(k) is the transfer function describing the evolution of the density contrast. We adopt the approximation in 
\cite{Bardeen86} (see Appendix B). See also \cite{Eisenstein97a,Eisenstein97b} for further discussions on transfer functions for large scale structure.
At early times, where non-linear evolution may be neglected, the linear density contrast can be written in terms of an effective transfer kernel:
\be
 \delta ({\bf{k}},z) = \frac{M(k) \Phi_k}{(1 + z)} = D(z) M(k) \Phi_k; \quad
 M(k) = \frac{2}{3} \frac{k^2 T(k)}{\Omega_M H^2_0} .
\ee
Here, $D(z)$ is the linear growth factor normalised such that $D(z) \rightarrow 1 / (1 + z)$. This allows us to express 
the linear power spectrum in terms of the transfer kernel $M(k)$ and the primordial power spectrum $P_{\phi} (k)$:
\begin{equation}
 P_{\delta}(k,z) = \frac{M^2(k)}{(1 + z)^2} P_{\phi} (k) .
\end{equation}
\n
We can relate the bispectrum for the primordial density perturbations to the bispectrum for the primordial potential perturbations by: 
\begin{equation}
 B^{prim}_{\delta} (k_1 , k_2 , k_{12} , z) = D^3(z) M(k_1) M(k_2) M(k_{12}) B^{prim}_{\Phi} (k_1,k_2,k_{12},z) .
\end{equation}
Therefore, in the linear regime (i.e. valid at large length scales), the primordial
bispectrum for the local model $B^{prim}$ will evolve according to the following expression:
\begin{equation}
B\ad^{loc}(\bk_1 \bk_1,\bk_3; z) = 2f_{NL}^{loc} D^3(z)\left [ {{M}({k_3})
\over {M}({k_1}) {M}({k_2})} P\ad(k_1)P_\ad(k_2) + \rm{cyc.} \right ].
\label{eq:loc}
\end{equation}
\n
The primordial potential bispectrum for the equilateral type can be expressed as \citep{Creminelli06}:
\begin{equation}
 B^{equi}_{\phi} = 6 f^{equi}_{\rm{NL}} \left[ - (P_{\Phi} (k_1) P_{\Phi} (k_2) + \rm{cyc}. ) - 2 (P_{\Phi} (k_1) P_{\Phi} (k_2) P_{\Phi} (k_3))^{2/3} + 
(P_{\Phi} (k_1) P^2_{\Phi} (k_2) P^3_{\Phi} (k_3) + \rm{cyc.} )^{1/3} \right] .
\end{equation}
\n
The primordial density bispectrum will therefore evolve according to:
\begin{eqnarray}
&& B\ad^{equi} (\bk_1 \bk_1,\bk_3; z) =  6f_{NL}^{equi} D^3(z) \bigg[ - \left(
{{M}({k_3}) \over {M}({k_1}) {M}({k_2}}) P\ad(k_1)P_\ad(k_2) + \rm{cyc. perm.} \right ) \nonumber \\ 
&& -2 \left ( {{M}({k_1}) {M}({k_2}) {M}({k_3})} \right )^{1/3} \{
{P\ad(k_1)P\ad(k_2)P\ad(k_3)} \}^{2/3} + \left ({{M}_{k_1}^{1/3} \over
{M}_{k_2}^{2/3} {M}_{k_3}} \{ {{P\ad(k_1)P\ad(k_2)^2P\ad(k_3)^3}
\}^{1/3}+
\rm{cyc. perm.}}  \right ) \bigg] .
\label{eq:bi_equi}
\end{eqnarray}
\n
Unlike the local model, the functional form for the equilateral model 
does not hold any relationship to fundamental physics but should be interpreted as something akin to a fitting function that
describes a number of models for which the exact analytical expressions are more complicated.\\
\n
The folded configuration is maximised when two of the sides of the triangle are approximately equal such that $k_2 \approx k_3 \approx k1/2$. The folded 
bispectrum is approximated by the following functional form:

\begin{equation}
 B^{fold}_{\phi} = 6 f^{fold}_{\rm{NL}} \Bigg[ ( P_{\Phi}(k_1) P_{\Phi}(k_2) + \rm{cyc}. ) + 3 ( P_{\Phi} (k_1) P_{\Phi} (k_2) P_{\Phi} (k_3) )^{2/3} - (P_{\Phi} (k_1) P^2_{\Phi} (k_2) P^3_{\Phi} (k_3) + \rm{cyc}. )^{1/3} \Bigg] .
\end{equation}
\n
Following the outlined procedure we can re-write this in terms of the primordial density perturbations:
\begin{align}
B^{fold}_{\delta} (k_1 , k_2 , k_3 , z) &= \frac{6 f^{fold}_{\rm{NL}}}{D(z)} \Bigg[ \Bigg( \frac{M(k_3)}{M(k_1) M(k_2)} P_{\delta}(k_1 , z) 
P_{\delta}(k_2 , z) + cyc. \Bigg) \nonumber \\
&  + 3 ( M(k_1) M(k_2) M(k_3) )^{-1/3} \lbrace P_{\delta}(k_1 , z) P^{\delta}(k_2 , z)
P_{\delta}(k_3 , z) \rbrace^{2/3} \nonumber \\
&- \Bigg( \frac{M^{1/3}(k_1)}{M^{1/3}(k_2) M(k_3)} \lbrace P_{\delta}(k_1 , z) 
P_{\delta}(k_2,z)^2 P_{\delta}(k_3,z)^3 \rbrace^{1/3} + cyc. \Bigg) \Bigg] .
\label{eq:bi_fold}
\end{align}
\n
The evolution of the primordial bispectrum is different compared to the one generated by gravitational evolution. 
The primordial bispectrum demonstrate momenta dependence and as a result the shape of the bispectrum configuration differs between the various primordial models as well as that for gravitational instability. At large angular scales, which will be probed by future 
weak lensing surveys, the gravitational instability may not have erased the memory of primordial 
non-Gaussianity, which can provide complementary information to results obtained from
CMB surveys such as Planck \citep{PC06}.
\n
%
\section{Three-dimensional Density Fields}
\label{sec:3Dden}
\subsection{Gravitationally Induced Bispectrum}
The first scenario that we will investigate is a three-dimensional smoothed field with galaxy density contrast $\delta_g$. The matter bispectrum is given by 
Eq.(\ref{eqn:MatterBispectrum}) which, when substituted into the skewness parameters Eq.(\ref{eqn:SkewnessParameters}) in the limit $q \ll 1$, gives:
\begin{eqnarray}
  && S_0 (k_2,z) = \frac{1}{\sigma^4_0} \frac{1}{(2 \pi^2)} \int_0^{\infty} \frac{k^2_1 d k_1}{(2 \pi^2)} \int^{+1}_{-1} d \mu W(k_1 R) W(k_3 R) 
\left[ \frac{15}{7} + \frac{3}{2} \mu \left( \frac{k_1}{k_2} + \frac{k_2}{k_1} \right) + \frac{6}{7} \mu^2 \right] P_{\delta}(k_1) P_{\delta}(k_2) .
\end{eqnarray}

\n
We now assume a Gaussian window function $W(x) = \exp (-x^2 / 2)$ allowing us to make the following simplification \citep{Matsubara02}:
\begin{equation}
\label{eq:WindowFunc}
\frac{W \left(\sqrt{l^2_1 + l^2_2 + 2 l_1 l_2 \mu} \right) }{W(l_1) W(l_2)} = \exp(-l_1 l_2 \mu); \quad l = kR,
\end{equation}
with $R$ the smoothing scale and $k$ the Fourier mode.

The angular integration over $\mu$ can now be analytically performed by relating the terms in the above skewness parameter to terms corresponding to the 
$m^{\rm{th}}$ Legendre polynomial. This allows us to use the following relation between integrals over the Legendre polynomials
and the modified Bessel functions $I_\mu(z)$ \citep{Lokas95,Matsubara02}:
\begin{equation}
 \int^{1}_{-1} \; d \mu \; P_m (\mu) e^{- \mu z} = (-1)^m \sqrt{\frac{2 \pi}{z}} I_{m + 1/2} (z) .
\end{equation}
\n
The relevant Legendre polynomials are:
\begin{equation}
 P_0 (z) = 1 ; \qquad
 P_1 (z) = z ; \qquad
 P_2 (z) = \frac{1}{2} (3 z^2 - 1) ; \qquad
 P_3 (z) = \frac{1}{2} (5 z^3 - 3 z) ; \qquad
 P_4 (z) = \frac{1}{8} (35 z^4 - 30 z^2 + 3) .
\end{equation}
\n
The resulting skewness parameters are:
\ben
S_{0} (k_2,z) &=& \frac{1}{\sigma^4_0} \frac{1}{( 2 \pi^2 )} \int_0^{\infty} \frac{ l^2_1 d l_1}{( 2 \pi^2) R^3} W^2 (l_1) W (l_2) 
P_{\delta} \left(\frac{l_1}{R} \right) P_{\delta} \left(\frac{l_2}{R} \right) \sqrt{\frac{2 \pi}{l_1 l_2}} \nonumber \\
&& \times\left[ \frac{17}{7} I_{1/2} (l_1 l_2) - \frac{3}{2} \left( \frac{l_1}{l_2} + \frac{l_2}{l_1} \right) I_{3/2} (l_1 l_2) + \frac{4}{7} I_{5/2} (l_1 l_2) \right]
\label{eq:gi3ds0},\\
S_{1} (k_2,z) &=&  \frac{1}{\sigma^2_0 \sigma^2_1} \frac{1}{(2 \pi^2)} \int_0^{\infty} \frac{l^2_1 d l_1}{(2 \pi^2) R^5} W^2 (l_1) W (l_2)
P_{\delta} \left(\frac{l_1}{R} \right) P_{\delta} \left( \frac{l_2}{R} \right) \sqrt{ 2 \pi l_1 l_2} \nonumber \\ 
&&\times 
\left[ \frac{41}{28} \left( \frac{l_1}{l_2} + \frac{l_2}{l_1} \right) I_{1/2} ( l_1 l_2 ) - 
\left( \frac{99}{35} + \frac{3}{4}\left[\frac{l^2_1}{l^2_2} + \frac{l^2_2}{l^2_1}\right] \right) I_{3/2} (l_1 l_2) + 
\frac{11}{14} \left( \frac{l_1}{l_2} + \frac{l_2}{l_1} \right) I_{5/2} (l_1 l_2) - \frac{6}{35} I_{7/2} ( l_1 l_2 ) \right]\label{eq:gi3ds1}, \\
S_{2} (k_2,z) &=&  \frac{1}{\sigma^4_0} \frac{1}{(2 \pi^2)} \int_0^{\infty} \frac{l^2_1 d l_1}{(2 \pi^2) R^7}  W^2 (l_1) W(l_2) 
P_{\delta} \left( \frac{l_1}{R} \right) P_{\delta} \left( \frac{l_2}{R} \right) \sqrt{2 \pi} (l_1 l_2)^{3/2} \nonumber \\
&& \times 
\Bigg[ \frac{81}{35} I_{1/2} (l_1 l_2) - \frac{9}{10} 
\left( \frac{l_1}{l_2} + \frac{l_2}{l_1} \right) I_{3/2} (l_1 l_2) 
- \frac{99}{49} I_{5/2} (l_1 l_2) 
+ \frac{9}{10} \left( \frac{l_1}{l_2} + \frac{l_2}{l_1} \right) I_{7/2} (l_1 l_2) - \frac{72}{245} I_{9/2} (l_1 l_2) \Bigg] . \label{eq:gi3ds2}
\een

\n
where we have used the relationship $l = k R$ as introduced in Eq.(\ref{eq:WindowFunc}). \\

\begin{figure}
\begin{center}
{\epsfxsize=15 cm \epsfysize=5.2cm {\epsfbox[31 507 588 713]{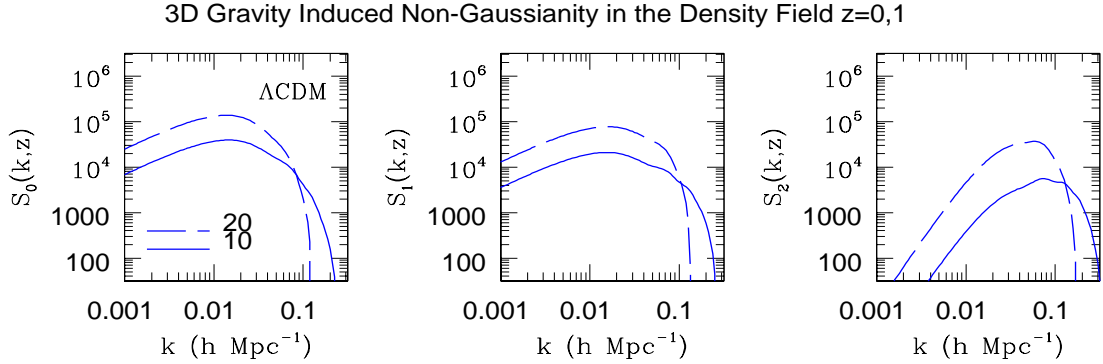}}}
\end{center}
\caption{The three {\em gravity induced} skew-spectra $S_0(k,z)$ (left panel), $S_1(k,z)$ (middle panel) and $S_2(k,z)$ (right panel) 
are depicted 3D cosmological density field. The spectra have been calculated for three different Gaussian 
smoothing scales $R=10,20{\rm\;h}^{-1}{\rm Mpc}$ for $z=0$. The skew-spectra for the redshifts $z=0$ and $z=1$ are
identical in the quasilinear regime due to the specific normalisation adopted.  
A $\Lambda$CDM background cosmology was assumed. The results are plotted for redshift $z=0$. The skew-spectra are normalized in such a 
way that they are virtually independent of redshift. The expressions for $S_0(k,z)$, $S_1(k,z)$ and  $S_2(k,z)$ are given in 
Eq.(\ref{eq:gi3ds0}), Eq.(\ref{eq:gi3ds1}) and Eq.(\ref{eq:gi3ds2}).}
\label{fig:3D-GI}
\end{figure}

\n
Using the notation for the skewness parameters proposed by Matsubara we can simplify the above equations with following variable:

\begin{equation}
S^{\alpha \beta}_{m} (l_2,R) = \frac{\sqrt{2 \pi}}{\sigma^4_0} \frac{1}{2 \pi^2} 
\left( \frac{\sigma_0}{\sigma_1 R} \right)^{\alpha + \beta - 2} \int \frac{l^2_1 dl_1}{2 \pi^2 R^3} 
P_{\delta} \left( \frac{l_1}{R} \right) P_{\delta} \left( \frac{l_2}{R} \right) e^{-l^2_1} e^{-l^2_2 / 2} l^{\alpha - 3/2}_1 l^{\beta - 3/2}_2 I_{m+1/2} 
(l_1 \, l_2).
\end{equation}
\n
In this notation the skew-spectra parameters become:
\begin{align}
& S_0 = \frac{17}{7} S^{11}_0 - 3 S^{(02)}_1 + \frac{4}{7} S^{11}_2 , \\
& S_1 = \frac{41}{14} S^{(13)}_0 - \frac{3}{2} S^{(04)}_1 - \frac{99}{35} S^{22}_1 + \frac{11}{7} S^{(13)}_2 - \frac{6}{35} S^{22}_3 , \\
& S_2 = \frac{81}{35} S^{33}_0 - \frac{9}{5} S^{(15)}_1 - \frac{99}{49} S^{33}_2 + \frac{9}{5} S^{(15)}_3 - \frac{72}{245} S^{33}_4 .
\end{align}
where we have adopted the following convention: $S^{(\alpha \beta)}_m = \frac{1}{2} ( S^{\alpha \beta}_m + S^{\beta \alpha}_m )$.
\n 
The results shown here demonstrate the analytic dependence of the gravitationally induced bispectra on a weighted integral over the modes in terms of Bessel functions. These equations are analogous to those presented in \cite{Matsubara02} but generalised to the skew-spectra formalism. The skew-spectra have been numerically calculated for three different smoothing scales $R[h^{-1} \rm{Mpc}] \in \lbrace 10,15,20 \rbrace$ and are shown in Figure \ref{fig:3D-GI}. The non-Gaussianity generated from gravitational instability gives rise to a positively skewed perturbation unlike a generic inflationary model which could predict negatively skewed perturbations (e.g. \cite{Hikage06}). It can also be seen that as the smoothing scale increases the amplitude increase but the scale at which we observe a cut-off increases (i.e. a cut-off occurs at smaller momenta). In the full skewness the gravitationally induced contributions are independent of smoothing scale \citep{Hikage06}. Additionally, due to the dependence of the gravitationally induced skew-spectra on the matter power spectrum and the variance of the field, we find that the generalised skew parameters are independent of redshift under the adopted normalisation.

\begin{figure}
\begin{center}
{\epsfxsize=15 cm \epsfysize=5.2cm {\epsfbox[31 507 588 713]{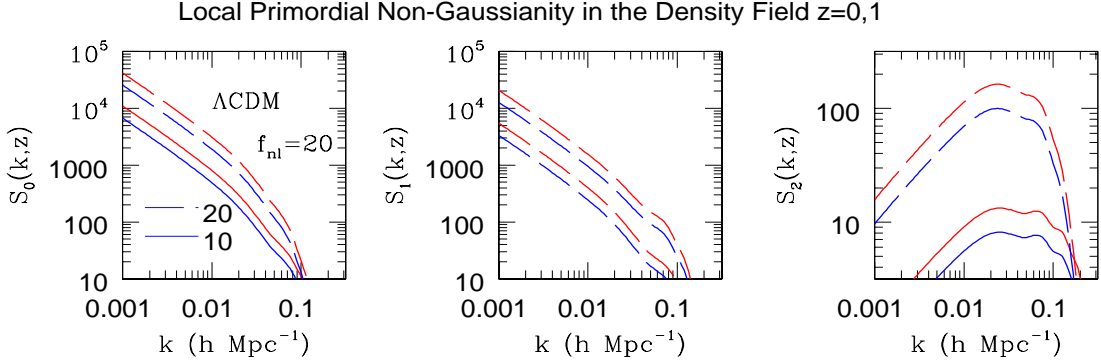}}}
\end{center}
\caption{The skew-spectra for primordial non-Gaussianity in a 3D cosmological density field is displayed for {\em local} type primordial non-Gaussianity. 
The spectra have been calculated for two different smoothing scales $R =10,20 {\rm h}^{-1}{\rm Mpc}$ with the left panel showing $S_0(k,z)$, the central panel $S_1(k,z)$ and the right panel $S_2(k,z)$. 
For each smoothing scale the upper curves correspond to higher redshift i.e. $z=1$
and the lower curves correspond to $z=0$.
The local type primordial non-Gaussianity is considered which is defined in Eq.(\ref{eq:bi_loc}). We have taken a $\Lambda$CDM cosmology with $f_{NL}=20$.}
\label{fig:3D-Loc}
\end{figure}
\begin{figure}
\begin{center}
{\epsfxsize=15 cm \epsfysize=5.2cm {\epsfbox[31 507 588 713]{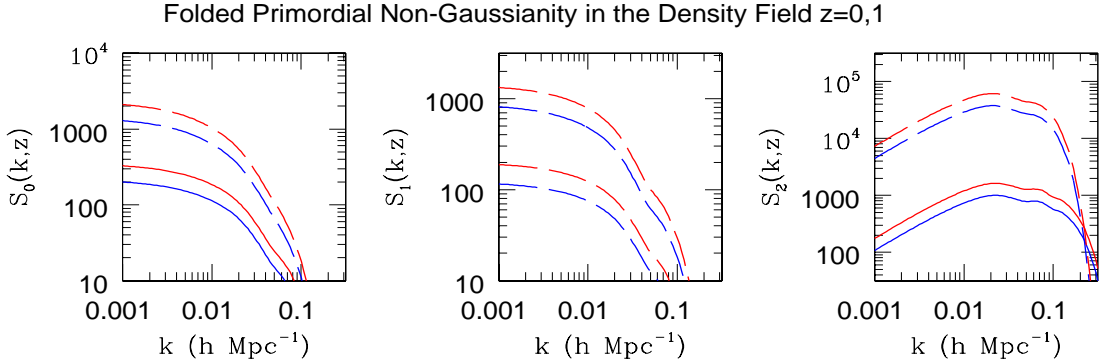}}}
\end{center}
\caption{Same as previous figure but for primordial non-Gaussianity of {\em folded} type Eq.(\ref{eq:bi_fold}).}
\label{fig:3D-Eq}
\end{figure}
\begin{figure}
\begin{center}
{\epsfxsize=15 cm \epsfysize=5.2cm {\epsfbox[31 507 588 713]{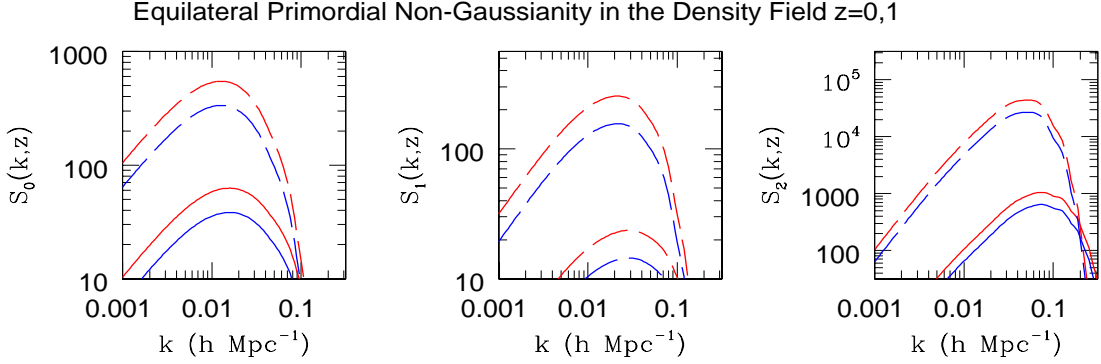}}}
\end{center}
\caption{Same as previous figure but for primordial non-Gaussianity of {\em equilateral} type Eq.(\ref{eq:bi_equi})}
\label{fig:3D-Fol}
\end{figure}
\begin{figure}
\begin{center}
{\epsfxsize=15 cm \epsfysize=5.2cm {\epsfbox[31 507 588 713]{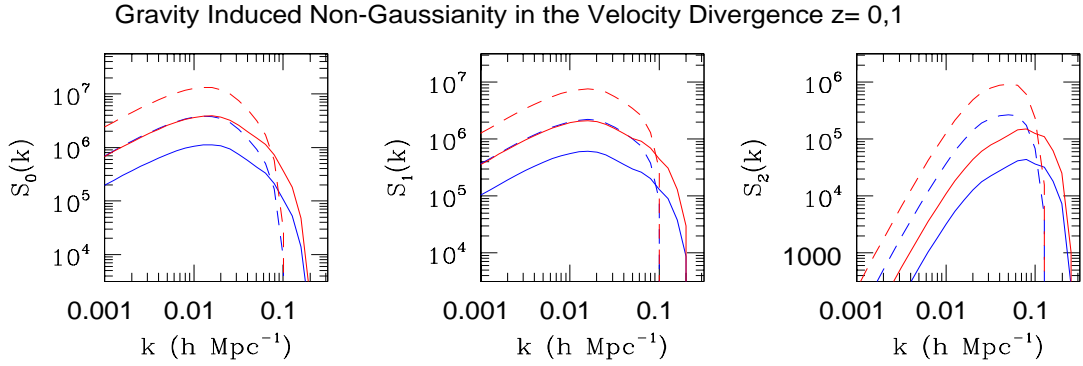}}}
\end{center}
\caption{The skew-spectra for {\em gravitational instability} in a 3D cosmological density field $\Theta$. The spectra have been calculated for three different smoothing scales $R =10,20 {\rm h}^{-1}{\rm Mpc}$ with the left panel showing $S_0(k,z)$, the central panel $S_1(k,z)$ and the right panel $S_2(k,z)$. 
The results for two different redshift slices are shown. For each smoothing scale the upper curves correspond to higher redshift i.e. $z_s=1$
and the lower curves correspond to $z_s=0$. The skew spectra are defined in Eq.(\ref{eq:ths0}), Eq.(\ref{eq:ths1}) and Eq.(\ref{eq:ths2}) respectively. The velocity divergence 
$\Theta$ is defined in Eq.(\ref{eq:theta_def}). The biaspectrum for $\Theta$ is given in Eq.(\ref{eq:ps_bps}). The redshift dependence for velocity divergence bispectrum depends of $g_{\theta}$ defined in Eq.(\ref{eq:g_theta}).}
\label{fig:Vel-GI}
\end{figure}
\begin{figure}
\begin{center}
{\epsfxsize=15 cm \epsfysize=5.2cm {\epsfbox[31 507 588 713]{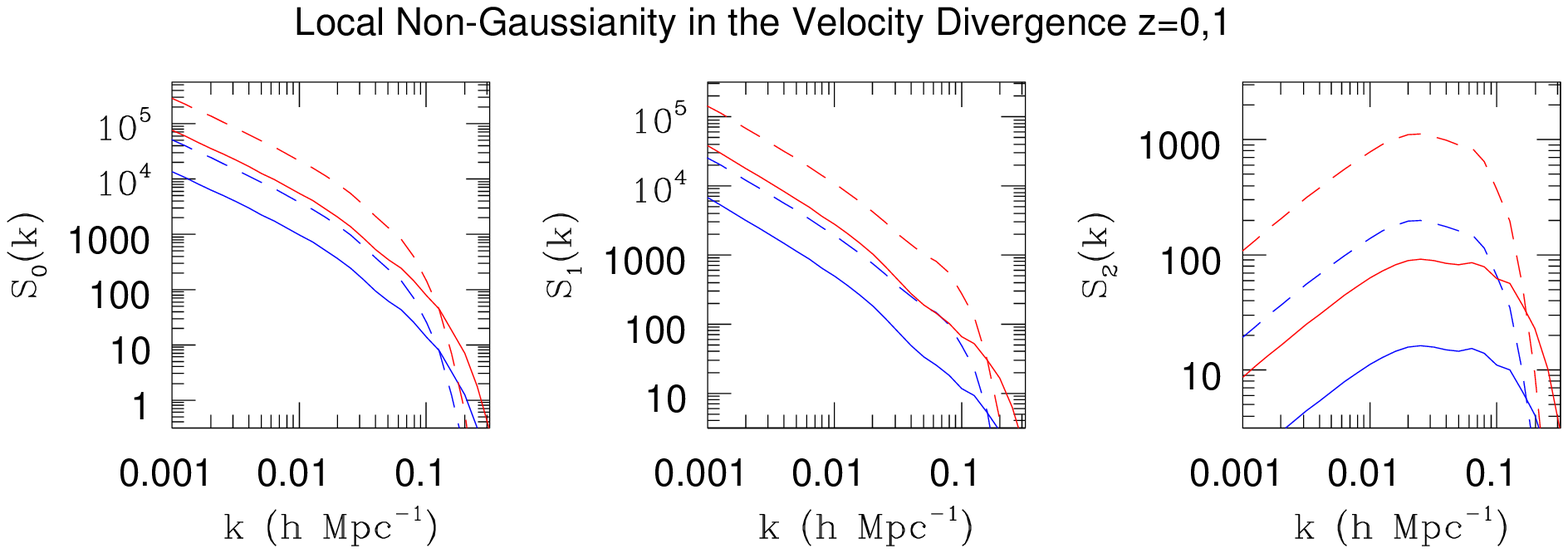}}}
\end{center}
\caption{Same as previous figure but for non-Gaussianity induced due to primordial non-Gaussianity of {\em local} type with $f_{NL}=20$.}
\label{fig:Vel-Loc}
\end{figure}
\begin{figure}
\begin{center}
{\epsfxsize=15 cm \epsfysize=5.2cm {\epsfbox[31 507 588 713]{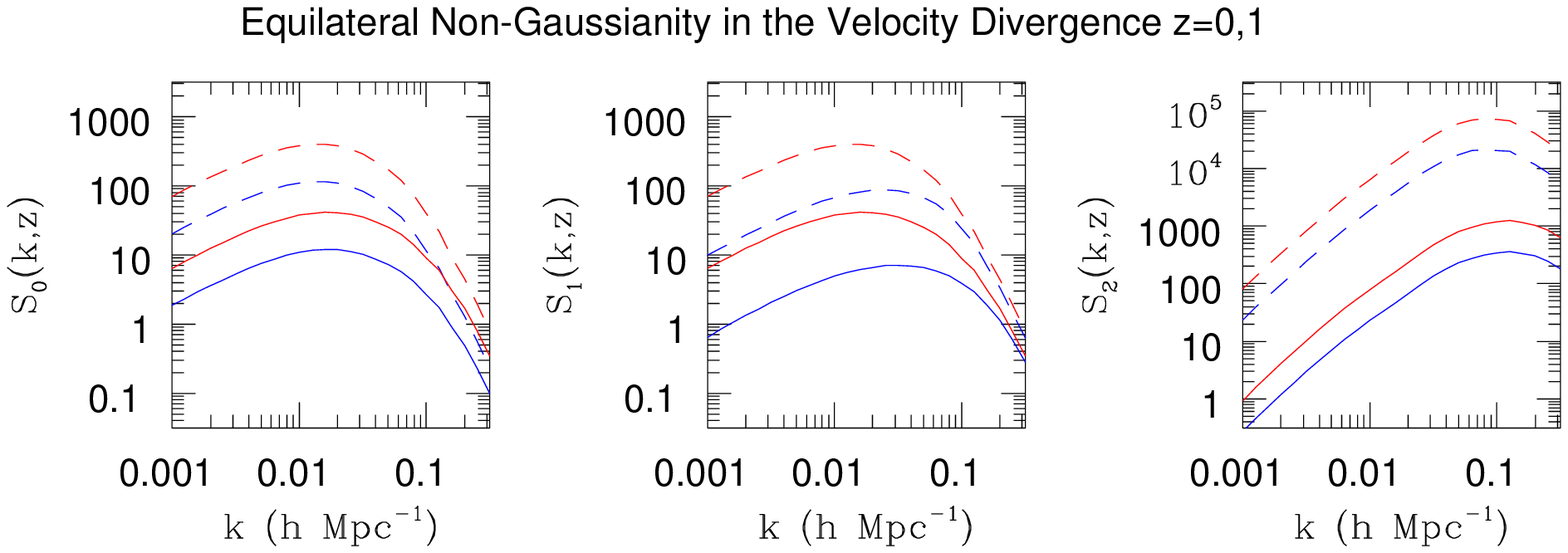}}}
\end{center}
\caption{Same as previous figure but for non-Gaussianity induced due to primordial non-Gaussianity of {\em equilateral} type with $f_{NL}=20$.}
\label{fig:Vel-Equ}
\end{figure}
\begin{figure}
\begin{center}
{\epsfxsize=15 cm \epsfysize=5.2cm {\epsfbox[31 507 588 713]{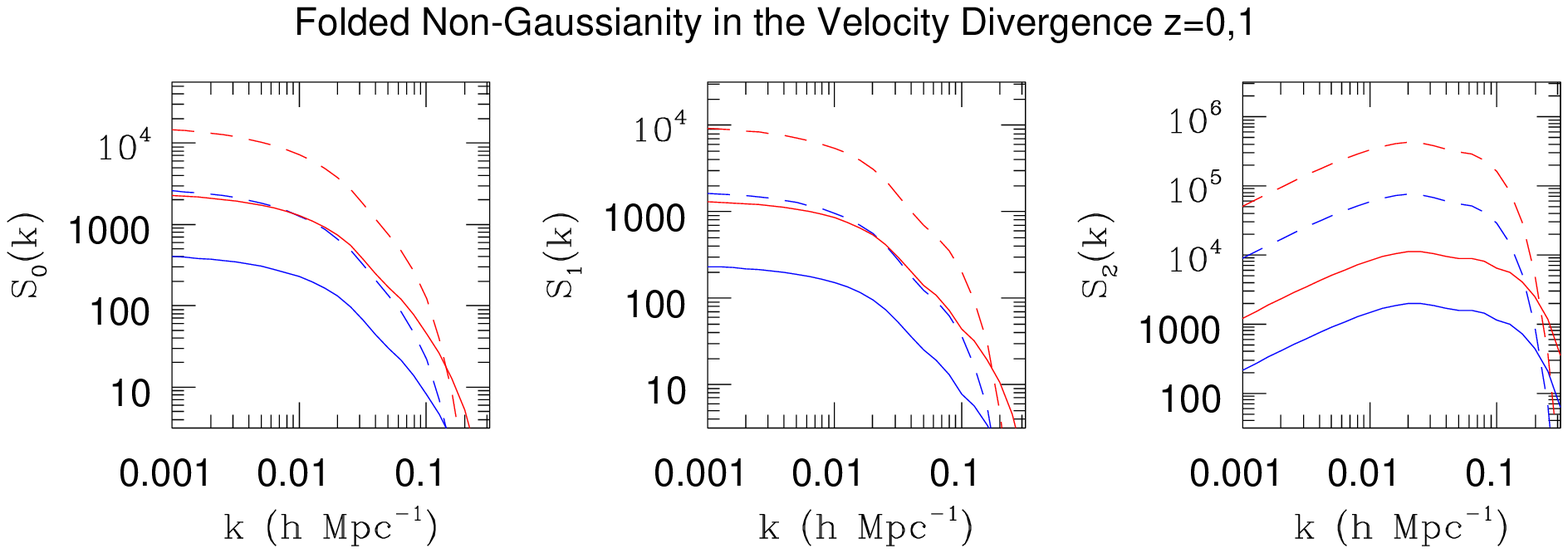}}}
\end{center}
\caption{Same as previous figure but for non-Gaussianity induced due to primordial non-Gaussianity of {\em folded} type with $f_{NL}=20$.}
\label{fig:Vel-Fol}
\end{figure}
\subsection{Primordial Bispectrum}
Using the above formalism we can calculate the skew-spectra for the primordial bispectrum in the local, equilateral and folded models. This is done by substituting 
the appropriate primordial bispectrum [Eqs.(\ref{eq:bi_loc},\ref{eq:bi_equi},\ref{eq:bi_fold})] into the expressions for the skew-spectra [Eqs.(\ref{eq:S0},\ref{eq:S1},\ref{eq:S2})] and evaluating. Due to the dependence of the primordial bispectra on the angular terms and the transfer function there does not appear to be a trivial analytical solution to these models and it is more elucidating to proceed by numerically integrating the resulting expressions allowing us to obtain the shape and amplitude of the spectra as a function of momenta. The resulting skew-spectra are shown in Figures (\ref{fig:3D-Loc},\ref{fig:3D-Eq},\ref{fig:3D-Fol}). Unlike gravitational instability the primordial models have redshift dependence and exhibit similar behaviour to the gravitationally induced models for varying smoothing scales. In the full skewness the primordial models grow in amplitude as we increase the smoothing scale \citep{Hikage06}. 
\section{Three-dimensional Velocity Field}
\label{sec:3Dvel}
The linear density perturbations $\delta_g$ was introduced in Eq.(\ref{eq:bi_loc}) as being related to Bardeen's curvature 
potential in the matter dominated era $\Phi$. In the Newtonian limit the Universe can be treated as a pressure-less fluid and the continuity equation 
allows us to relate the linear density perturbations to an associated peculiar velocity ${\bf{v}} ( {\bf{k}} , z )$:
\begin{equation}
 \dot{\delta_g} ( {\bf{k}} ) + \nabla \cdot {\bf{v}} ( {\bf{k}} ) = 0 .
\end{equation}
In the basic inflationary paradigm and observationally the linear primordial density perturbations are highly Gaussian. Through the continuity equation 
we expect that deviations from Gaussianity will impact the observed peculiar velocity field. Following \cite{Matsubara02} we introduce a 
dimensionless scalar field normalised by the Hubble variable $H$:
\begin{equation}
\Theta({\bf{x}}) = {\rm H^{-1}} \nabla \cdot {\bf{v}} ( {\bf{x}} ).
\label{eq:theta_def}
\end{equation}
Due to the expansion of the Universe, the velocity field is described purely by the diverge (i.e. rotation-free) term with the rotational components being 
described by decaying solutions in perturbation theory (see, for example, \cite{Matsubara02,Bernardeau02}). Using second order perturbation theory the 
power spectrum and bispectrum can be analytically calculated (\cite{Bernardeau94,Matsubara02}):
\ben
&& P_{\Theta}(k) = g^2_{\theta} P_{\delta} (k) + \mathcal{O} (\sigma^4_0); \\
&& B_{\Theta}(k_1, k_2, k_3) = -g^3_{\theta} \left[ \frac{6}{7} + \mu \left(\frac{k_1}{k_2} + \frac{k_2}{k_1} \right) + \frac{8}{7} \mu^2 \right] 
P_{\delta}(k_1)P_{\delta}(k_2) + cyc.perm. + \mathcal{O} (\sigma^6_0).
\label{eq:ps_bps}
\een
The factor $g_{\theta}$ is the logarithmic derivative of the growth factor $D(z)$ with respect to the scale factor $a$:
\be
g_{\theta}={d \ln D \over d \ln a} = \Omega_M^{4/7} + {\Lambda \over 70}\left ( 1+ {\Omega_M \over 2} \right ) .
\label{eq:g_theta}
\ee 
\n
\subsection{Gravitationally Induced Bispectrum}
Using the modified power spectrum and bispectrum Eq.(\ref{eq:ps_bps}) we can calculate the skewness parameters for the peculiar velocity field:
\ben
S_{0} (l_2,z) &=& \frac{-1}{g_{\theta} \sigma^4_0} \frac{1}{2 \pi^2} \int \frac{l^2_1 dl_1}{2 \pi^2 R^3} W^2(l_1) W(l_2) P_{\delta} \left( \frac{l_1}{R} \right)
P_{\delta} \left( \frac{l_2}{R} \right) \sqrt{\frac{2 \pi}{l_1 l_2}} \nn \\
&& \times \Bigg[ \frac{13}{7} I_{1/2} (l_1 l_2) - \frac{3}{2}\left( \frac{l_2}{l_1} + 
\frac{l_1}{l_2} \right) I_{3/2} (l_1 l_2) + \frac{8}{7} I_{5/2} (l_1 l_2) \Bigg] , \label{eq:ths0}\\
S_{1} (l_2,z) &=&  \frac{-1}{g_{\theta} \sigma^2_0 \sigma^2_1} \frac{1}{2 \pi^2} \int \frac{l^2_1 dl_1}{2 \pi^2 R^5} W^2(l_1) W(l_2) P_{\delta} 
\left( \frac{l_1}{R} \right) P_{\delta} \left( \frac{l_2}{R} \right) \sqrt{2 \pi l_1 l_2} \nn \\
&&  \times 
\Bigg[\frac{33}{28} \left( \frac{l_1}{l_2} + \frac{l_2}{l_1} \right) I_{1/2} (l_1 l_2) - \left[ \frac{3}{4} \left( \frac{l^2_1}{l^2_2} + \frac{l^2_2}{l^2_1} \right) + \frac{93}{35} \right] I_{3/2} (l_1 l_2) + \frac{15}{14} \left( \frac{l_1}{l_2} + \frac{l_2}{l_1} \right) I_{5/2} (l_1 l_2) - \frac{12}{35} I_{7/2} (l_1 l_2)  \Bigg] ,\label{eq:ths1} \\
S_{2} (l_2,z) &=&  \frac{-1}{g_{\theta} \sigma^4_1} \frac{1}{2 \pi^2} \int \frac{l^2_1 dl_1}{2 \pi^2 R^7} W^2(l_1) W(l_2) P_{\delta} \left( \frac{l_1}{R} \right) 
P_{\delta} \left( \frac{l_1}{R} \right) \sqrt{2 \pi} (l_1 l_2)^{3/2} \nn \\ \nonumber&& \\ \nonumber&& \\
&& \times \Bigg[
\frac{57}{35} I_{1/2} (l_1 l_2) - \frac{9}{10} \left( \frac{l^3_1}{l_2} + \frac{l^3_2}{l_1} \right) I_{3/2} (l_1 l_2) - \frac{51}{49} I_{5/2} (l_1 l_2) + \frac{9}{10} \left( \frac{l^3_1}{l_2} + \frac{l^3_2}{l_1} \right) I_{7/2} (l_1 l_2) - \frac{144}{245} I_{9/2} (l_1 l_2)
\Bigg] . \label{eq:ths2} 
\een

Using the parameterisation presented for the 3-dimensional cosmological density fields we can simplify the above expressions:

\begin{align}
 S_0 (k_2,R) &= \frac{1}{g_{\theta}} \left( - \frac{13}{7} S^{11}_0 + \frac{3}{2} S^{(02)}_1 - \frac{8}{7} S^{11}_2 \right) , \\
 S_1 (k_2,R) &= \frac{1}{g_{\theta}} \left( - \frac{33}{14} S^{(13)}_0 + \frac{3}{2} S^{(04)}_1 - \frac{93}{35} S^{22}_1 - \frac{15}{7} S^{(13)}_2 + \frac{12}{35} S^{22}_3 \right), \\
 S_2 (k_2,R) &= \frac{1}{g_{\theta}} \left( - \frac{57}{35} S^{33}_0 + \frac{9}{5} S^{(15)}_1 + \frac{51}{49} S^{33}_2 - \frac{9}{5} S^{(15)}_3 + \frac{144}{245} S^{33}_4 \right) .
\end{align}

These are, again, analogous to the appropriate expressions in \cite{Matsubara02}. The skew-spectra for the three different smoothing scales are shown in 
figure \ref{fig:Vel-GI}.
\subsection{Primordial Bispectrum}
The 3D cosmological velocity field can be used to probe primordial contributions. We calculate the skew-spectra for the local model (Figure \ref{fig:Vel-Loc}), the equilateral model (Figure \ref{fig:Vel-Equ}) and the folded model (Figure \ref{fig:Vel-Fol}). All models have been calculated for redshifts $z = 0,1$ and for smoothing scales $R = 10,20 \, h^{-1} \rm{Mpc}$.
\section{Two-Dimensional Projected Density Fields}
\label{sec:2D}
Often cosmological data sets can be cast in the form of a projection over the sky. This requires us to introduce a correspondence between three-dimensional 
density fields and the observed two-dimensional projection. The formalism adopted here follows \cite{Matsubara02}. 
When working with a two-dimensional projected field one of the key observables are angular distances for objects placed at a given comoving distance $r$. 
Under the assumption of statistical homogeneity and isotropy the Universe can be approximated by the FRW cosmological models. The line element
for an FRW cosmology can be expressed as:
\begin{equation}
 ds^2 = -c^2dt^2 + a^2(t) \left[ dr^2 + d_A(r)^2(\sin^2\theta d\theta^2 + d\phi^2) \right].
\end{equation}
\n
where we $d_A(r)$ is the angular diameter distance to radial comoving distance $r$. The scale factor is denoted by $a(t)$.
We have $d_A(r)=\sinh({\rm K}^{-1/2}r), r, \sin({\rm K}^{-1/2}r)$ for a Universe of negative, zero and positive curvature respectively. The curvature is given by ${\rm K}=(\Omega_M-1){\rm H}_0^2$.
The three dimensional density field is then projected onto a two-sphere through the use of a selection function, $n(\br)$, and the angular diameter distance $d_A(r)$
to relate the projected density contrast field $\Psi(\oh)$ to the underlying three-dimensional density field $\delta_g$. Under the flat sky approximation this 
reduces to:
\begin{equation}
 \Psi(\oh) = \int\;dr\;d^2_{A}(r)\; \delta_g\left(\br,r\right) n (r) .
\end{equation}
\begin{figure}
\begin{center}
{\epsfxsize=15 cm \epsfysize=5.2cm {\epsfbox[31 507 588 713]{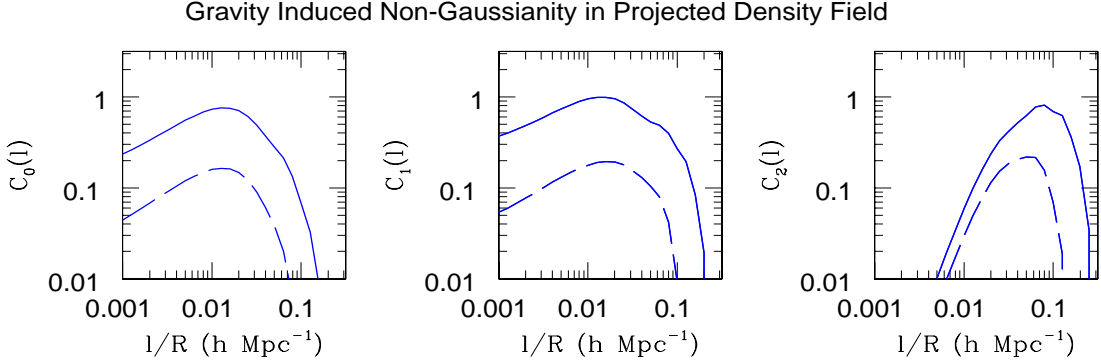}}}
\end{center}
\caption{The skew-spectra induced by gravitational instability for projected (2D) cosmological density field. 
The spectra have been calculated for two different smoothing angular scales $R = 10,20 h \rm{Mpc}^{-1}$ 
with the left panel showing $C_0 (l)$, the central panel $C_1 (l)$ and the right panel $C_2 (l)$. The solid line corresponds to $R = 10 h \rm{Mpc}^{-1}$ and the dashed line corresponds to $R = 20 h \rm{Mpc}^{-1}$. The projected skew-spectra are
defined in Eq.(\ref{eq:2ds0}), Eq.(\ref{eq:2ds1}) and Eq.(\ref{eq:2ds2}).}
\label{fig:2D-GI}
\end{figure}
\begin{figure}
\begin{center}
{\epsfxsize=15 cm \epsfysize=5.2cm {\epsfbox[31 507 588 713]{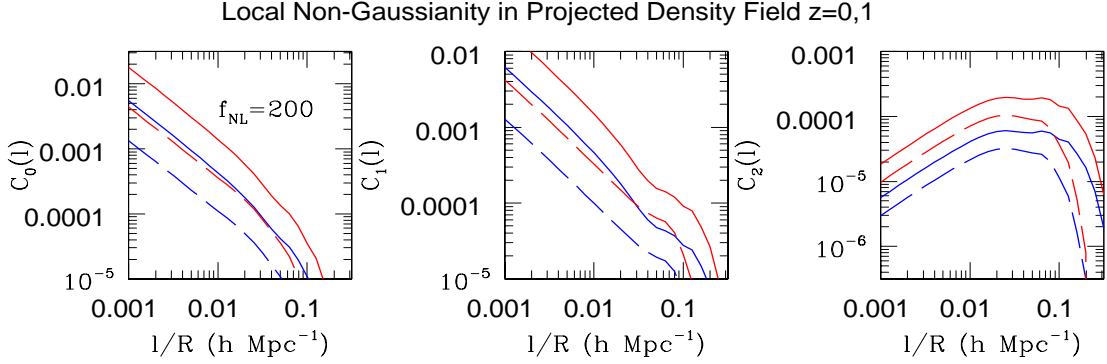}}}
\end{center}
\caption{Same as previous figure but for {\em local} type primordial non-Gaussianity for $f_{NL}=200$.}
\label{fig:2D-Loc}
\end{figure}
\begin{figure}
\begin{center}
{\epsfxsize=15 cm \epsfysize=5.2cm {\epsfbox[31 507 588 713]{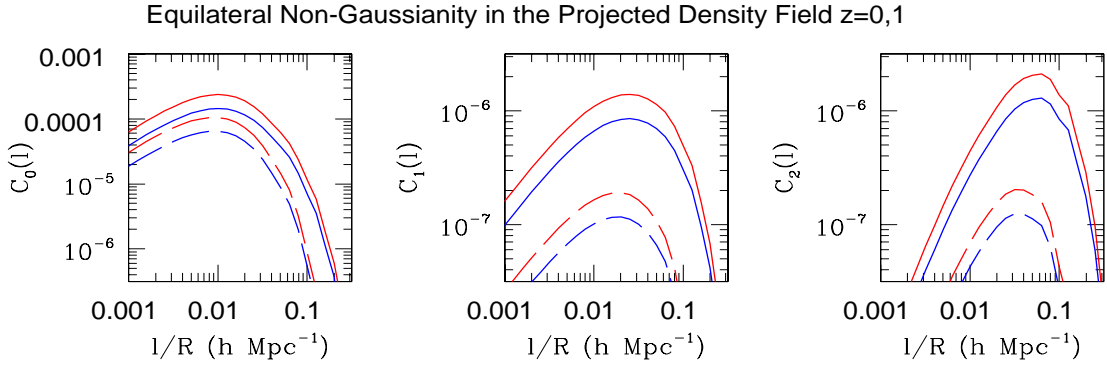}}}
\end{center}
\caption{Same as previous figure but for {\em equilateral} type primordial non-Gaussianity.}
\label{fig:2D-Fol}
\end{figure}
\begin{figure}
\begin{center}
{\epsfxsize=15 cm \epsfysize=5.2cm {\epsfbox[31 507 588 713]{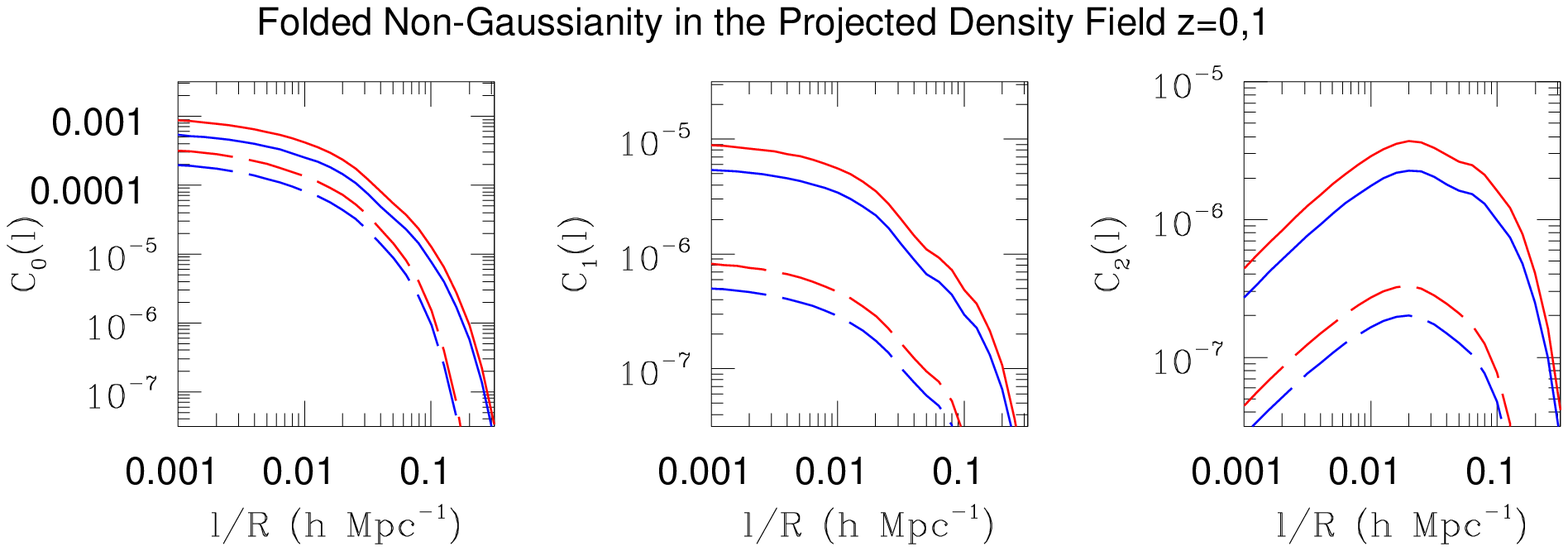}}}
\end{center}
\caption{Same as previous figure but for {\em folded} type primordial non-Gaussianity.}
\label{fig:2D-Equ}
\end{figure}
\n
We can use a similar approach to move beyond one point statistics and consider the projected polyspectra. The two lowest order spectra of interest to us are 
the projected power spectrum $P_{2D}$ and the projected bispectrum $B_{2D}$:
\begin{align}
 & P_{\Psi} (\ell) = \int dr \; d_A^2(r)\;n^2(r)P_{\delta} \left[ \frac{\ell}{d_{A}(r)} ; r \right]; \\
 & B_{\Psi} (\ell_1 , \ell_2 , \ell_3 ) = \int dr d^2_{A}(r) \;  n^3(r) B_{\delta} \left[ \frac{\ell_1}{d_{A}(r)} ; \frac{\ell_2}{d_{A} (r)} ;
 \frac{\ell_3}{d_{A}(r)} ; r \right] .
\end{align}
The three-dimensional power spectra $P_{\delta}(k;r)$ has already been specified and the bispectrum $B_{\delta}(k_i;r)$ will be evaluated 
for the appropriate model. In an analogous manner 
to the three-dimensional density fields we can construct variance and skewness parameters to allow us to investigate the contributions from various sources 
of non-Gaussianity to the projected bispectrum. The main difference in this approach is that we will be required to introduce the necessary formalism to 
cope with projection effects. The necessary machinery is discussed in \cite{Matsubara02} though we have introduced the appropriate 
modifications to generalise the results to the proposed skew-spectra approach.
\n
The variance parameters of a projected field smoothed by an arbitrary window function, $W(\ell\theta_b)$, at a given smoothing scale, $\theta_b$, is 
given by:
\begin{equation}
 \sigma^2_j \left( \theta_b \right) = \int \frac{\ell d \ell}{2 \pi} \ell^{2 j} P (\ell) W^2 (\ell\theta_b) = \frac{1}{\theta^{2j+1}_b} \int 
dr d^2_{A} (r) n^2(r) D^2 (r) \Sigma^2_j \left[ d_{A}(r) \theta_b \right] .
\end{equation}
We have introduced a variance function $\Sigma_j$ to relate the three-dimensional field to our projected field:
\begin{equation}
 \Sigma^2_j (R) = R^{2j + 2} \int \frac{k dk}{2 \pi} k^{2j} P^{lin}_{\delta} (k) W^2(kR) .
\end{equation}
The skew-spectra parameters can then be found by introducing the appropriate functions of the projected field and projected derivative fields 
Eq.(\ref{eqn:SkewnessParameters}):
\begin{equation}
 S^{j}_{2D} (\theta_b) = \frac{1}{\sigma^4_0 \theta^4_b} \left( \frac{\sigma_0}{\sigma_1 \theta_b} \right)^{2j} \int dr d^2_{A}(r) n^3 (r) D^4 (r) 
\Sigma^{4 - 2j}_0 \left[ d_{A}(r) \theta_b \right] \Sigma^{2j}_1 \left[ d_{\rm A}(r) \theta_b \right] \mathcal{C}^{(j)} \left[d_A(r) \theta_b \right] ,
\end{equation}
where skew-spectra parameters are given by:
\begin{align}
  & \mathcal{C}_0 (l_2,R) = \frac{6}{\Sigma^4_0} \int \frac{l_1 d l_1}{2 \pi} \int \frac{d \mu}{2 \pi \sqrt{1 - \mu^2}}B_{\delta}(l_1 , l_2 , l_3 , z ) W(l_1\theta_b) 
  W (| {\bf{l}}_1 + {\bf{l}}_2 |\theta_b) \label{eq:2ds0} , \\
  & \mathcal{C}_1 (l_2,R) = \frac{3}{\Sigma^2_0 \Sigma^2_1} \int \frac{l_1 d l_1}{2 \pi} \int \frac{d \mu}{2 \pi \sqrt{1 - \mu^2}} 
  |{\bf{l}}_1 + {\bf{l}}_2|^2 B_{\delta}(l_1 , l_2 , l_3 , z ) W(l_1\theta_b)
  W(| {\bf{l}}_1 + {\bf{l}}_2 |\theta_b) \label{eq:2ds1} , \\
  & \mathcal{C}_2 (l_2,R) = \frac{12}{\Sigma^4_1} \int \frac{l_1 d l_1}{2 \pi} \int \frac{d \mu}{2 \pi \sqrt{1 - \mu^2}} 
  ({\bf{l}}_1 \cdot {\bf{l}}_2) |{\bf{l}}_1 + {\bf{l}}_2|^2 B_{\delta}(l_1 , l_2 , l_3 , z ) W(l_1\theta_b) 
  W (| {\bf{l}}_1 + {\bf{l}}_2 |\theta_b) \label{eq:2ds2} .
\end{align}
%

For the 2-dimensional spectra the angular integration is performed by using the integral representation of the modified Bessel functions $I_m (l_1 l_2)$ along with the recursion relations $I^{\prime}_m = (I_{m-1} + I_{m+1}) / 2$. The integral representation is given by:

\begin{equation}
 \frac{1}{\pi} \int^{1}_{-1} \frac{ d \mu}{\sqrt{1 - \mu^2}} \mu^m e^{- l_1 l_2 \mu} = \left( - \frac{ d}{d (l_1 l_2)} \right)^m I_0 (l_1 l_2); \quad\quad  I_0 (l_1 l_2) = \frac{1}{\pi} \int^{1}_{-1} \frac{ d \mu}{\sqrt{1 - \mu^2}} e^{-l_1 l_2 \mu} .
\end{equation}

\n
Proceeding analogously to the 3-dimensional case the 2-dimensional skew-spectra reduce to a weighted integral over the the modes $k = l / R$:

\begin{align}
 C_0 (l_2,R) &= \frac{1}{\Sigma^4_0} \int \frac{l_1 dl_1}{2 \pi} P \left( \frac{l_1}{R} \right) P \left( \frac{l_2}{R} \right) W^2 (l_1) W(l_2) \Bigg[ 
\frac{23}{7} I_0 (l_1 l_2) - 3 \left( \frac{l_1}{l_2} + \frac{l_2}{l_1} \right) I_1 (l_1 l_2) + \frac{6}{7} I_2 (l_1 l_2) \Bigg] , \\
 C_1 (l_2,R) &= \frac{1}{\Sigma^2_0 \Sigma^2_1} \int \frac{l_1 dl_1}{2 \pi} P \left( \frac{l_1}{R} \right) P \left( \frac{l_2}{R} \right) W^2(l_1) W(l_2) \nonumber
\\ &\times \Bigg[ \frac{93}{28} (l^2_1 + l^2_2) I_0 (l_1 l_2) - \frac{3}{2} \left( \frac{l^3_1}{l_2} + \frac{l^3_2}{l_1} + \frac{27}{7} l_1 l_2 \right) I_1 (l_1 l_2) 
+ \frac{33}{28} (l^2_1 + l^2_2) I_2 (l_1 l_2) - \frac{3}{14} l_1 l_2 I_3 (l_1 l_2) \Bigg] , \\
 C_2 (l_2,R) &= \frac{1}{\Sigma^4_1} \int \frac{l_1 dl_1}{2 \pi} P \left( \frac{l_1}{R} \right) P \left( \frac{l_2}{R} \right) W^2(l_1) W(l_2) \nonumber \\
&\times \Bigg[ \frac{33}{7} l^2_1 l^2_2 I_0 (l_1 l_2) - \frac{3}{2} (l^3_1 l_2 + l^3_2 l_1) I_1 (l_1 l_2) - \frac{30}{7} l^2_1 l^2_2 I_2 (l_1 l_2) - \frac{3}{2} (l^3_1 l_2 + l^3_2 l_1) I_3 (l_1 l_2) + l^2_1 l^2_2 I_4 (l_1 l_2) \Bigg] .
\end{align}

\n
We adopt a parameterisation in terms of the variable $C^{\alpha \beta}_m$:

\begin{equation}
 C^{\alpha \beta}_m (l_2,R) = \frac{1}{\Sigma^4_0} \left( \frac{\Sigma_0}{\Sigma_1} \right)^{\alpha + \beta - 2} \int \frac{l_1 dl_1}{2 \pi}  P \left( \frac{l_1}{R} \right) P \left( \frac{l_2}{R} \right) W^2 (l_1) W(l_2) I_m (l_1 l_2) l^{\alpha - 1}_1 l^{\beta - 1}_2 .
\end{equation}

\n
The skew-spectra can then be written as:

\begin{align}
 C_0 (l_2,R) &= \frac{23}{7} C^{11}_0 - 6 C^{(02)}_1 + \frac{6}{7} C^{11}_2 , \\
 C_1 (l_2,R) &= \frac{93}{14} C^{(31)}_0 - 3 C^{(40)}_1 - \frac{81}{14} C^{22}_1 + \frac{33}{14} C^{((31)}_2 - \frac{3}{14} C^{22}_3 , \\
 C_2 (l_2,R) &= \frac{33}{7} C^{33}_0 - 3 C^{(42)}_1 - \frac{30}{7} C^{33}_2 - 3 C^{42}_3 + C^{33}_4 .
\end{align}

\n
The resulting skew-spectra for gravitational instability are shown in Figure (\ref{fig:2D-GI}) for smoothing scales $R = 10,20 \, h \rm{Mpc}^{-1}$. 

The 2D projected surveys can be used with the primordial bispectra by simply replacing the bispectrum kernel in 
Eqs.(\ref{eq:2ds0},\ref{eq:2ds1},\ref{eq:2ds2}) with the corresponding primordial model. The results are shown in Figures (\ref{fig:2D-Loc},\ref{fig:2D-Fol},\ref{fig:2D-Equ}).

\section{Real World Effects}
\label{sec:realwo}
In this section we briefly outline some key observational and systematic effects that lead to corrections and modifications to measured quantities 
(e.g. power spectrum). This paper does not attempt to deal with real world effects but a more general treatment incorporating real 
world effects will be presented elsewhere.

\subsection{Redshift Space Distortions}
Real world observations and surveys consist of constructing a map detailing the frequency and intensity of radiation across the sky or a patch of the sky. 
Cosmological distances are constructed under the assumption that the frequency of a particular emitter (e.g. CO emission) is known allowing us to 
estimate the redshift to the source. Under the assumption that we inhabit a spatially homogeneous, isotropic, irrotational and expanding spacetime then 
the redshift may be converted to a coordinate distance in a fairly straightforward manner. 
Redshift distance estimates consist of a number of contributions of which the two most 
physically relevant, for cosmological observations, are the metric expansion of spacetime (as the dominant factor) and the Doppler shifts arising from 
local peculiar velocities of a source along the line of sight.

The non-linear nature of the mapping from real space to redshift space means that a 
Gaussian random field will have non-Gaussian corrections when observed in redshift space. In addition to these non-Gaussian corrections there will be 
modifications to the observed power spectrum. Such systematics will become increasingly important with the precision of the surveys mentioned in 
Section \ref{sec:int}. There exists analytic results for the linear results (\cite{Kaiser87,Hamilton92,Hamilton97,Matsubara00}) and recent studies 
have begun to investigate the 
full non-linear redshift-space power spectra (e.g. \cite{Shaw08}). The linear result first provided in \cite{Kaiser87} introduces the following 
modification to the 2-point function: 

\begin{equation}
P_{g,s} (k) = b^2 (1 + b^{-1} f \mu^2_{k} )^2 P_{\delta} (k) ,
\end{equation}

\n
where b is a linear bias factor, f the derivative of the linear growth factor and $\mu_k = \hat{{\bf{n}}} \cdot \hat{{\bf{k}}}$.

\subsection{Galaxy Bias}
The formation of galaxies and collapsed objects is highly dependent on the total local matter density. 
An important point to note is that large scale structure surveys only probe the luminous baryonic matter and we find that there are nonlinearities 
in the biased relationship between the observable baryonic matter distribution (i.e. galaxies) and the underlying total matter distribution 
(i.e. baryonic plus cold dark matter). This nonlinearity can give rise to an effective non-Gaussianity that must be taken into account 
when performing statistical tests of large scale structure. Following \cite{Fry93} we can assume that galaxy formation is a local 
deterministic process and perturbatively expand the galaxy density contrast $\delta_g$ in terms of the underlying matter density contrast $\delta_m$:

\begin{equation}
 \delta_g (z) = b_0 (z) + b_1 (z) \delta_m (z) + \frac{b_2 (z)}{2} \delta^2_m (z) + \mathcal{O} ( \delta^3_m ) ,
\end{equation}

where $b_0 (z)$ is set by the constraint $\langle \delta_g (z) \rangle = 0$. The two remaining parameters 
$b_1 (z)$ and $b_2(z)$ are the galaxy bias parameters. The 2-point and 3-point correlators of the galaxy distribution are therefore weighted 
functions of the 2-point and 3-point correlators of the matter distribution:

\begin{align}
 P_g (k , z) &= b^2_1 (z) P_m (k,z) , \\
 B_g (k_1 , k_2 , k_3 , z) &= b^3_1(z) B_m (k_1 , k_2 , k_3 , z) + b^2_1 (z) b_2 (z) \left[ P_m (k_1 , z) P_m (k_2 , z) + \rm{cyc.} \right] .
\end{align}

In the above $B_m (k_1,k_2,k_3,z)$ is the bispectrum of the matter bispectrum. For this report we will neglect the bispectral contributions 
of galaxy biasing and focus primarily on the primordial non-Gaussianities, for a detailed discussion of biasing effects see 
\cite{Fry93,Scoccimarro00,McDonald06,McDonald08,Sefusatti07,Jeong09a,Jeong09b} for a selection of relevant papers.

\subsection{Survey Effects}
In addition to redshift space distortions and biasing effects we also have to consider additional observational effects such as survey geometry, 
survey volume and corrections to the radial selection function arising from magnitude limited surveys. Other effects include the survey 
specifications (e.g. range of redshifts probed), instrumental noise and additional complex non-linearities that can manifest at high k altering 
the power spectrum. Examples of these complex non-linearities include both instrumental transfer function effects in Fourier space 
and survey specific sample biasing. A specific model for the instrument transfer function must be assumed to investigate effects at high k for a given 
survey. Such a modification is arbitrary and survey dependent. Skew spectra, being high order statistics, will be affected more by
sample (cosmic) variance. The resulting bias and scatter of these estimators will have to be dealt with more precautions than their 
ordinary power spectrum counterparts. 
Such issues will be dealt with in a separate publication. 
We shall provide a quick discussion of magnitude limited surveys, radial selection function and volume of the survey following the discussion 
presented in \citep{Hikage02}.

In an apparent-magnitude limited survey the number density of observed galaxies will decrease with redshift. This means that for the 
galaxies to be appropriately resolved, and hence a reliable density field reconstructed, 
we need an appropriate smoothing length $R > R_g$ such that $R_g$ is the mean separation of galaxies at our 
specified redshift. This means that we have a maximum redshift $z_{\rm{max}} = z_{\rm{max}} (R_g)$ such that $R_g$ will be the mean separation of the magnitude 
limited galaxy sample at our maximum redshift $z_{\rm{max}}$ \citep{Vogeley94}. The number of independent resolution elements in each survey volume 
will be given by \citep{Hikage02} $N_{\rm{Res}} = V_{\rm{surv}} ( < Z_{\rm{max}}) / (2 \pi)^{3/2} R^3_{g}$. This paper does not attempt to take real-world 
systematics into account, a treatment of real-world systematics and error estimates will be presented elsewhere. 

\section{Conclusion}
\label{sec:conclu}
New surveys proposed over the next decade or two offer the possibility of probing large scale structure in an unprecedented manner. It is hoped that larger, more comprehensive cosmological data sets along with a more sophisticated understanding of the underlying systematics will allow us to investigate the role of non-Gaussian initial conditions along with the physics underlying gravitational instability in an FLRW Universe. This makes the statistics of galaxy clustering an interesting playground for model testing and discrimination. 

In this paper we have adopted the conventional approach to describing gravitational clustering through a hierarchy of higher order correlation functions. There exist numerous statistical methods to characterise the higher order correlators ranging from polyspectra defined in Fourier space to the spatially defined Minkowski Functionals adopted in this paper. The MFs are a topological statistic that characterise the morphological properties of a random field. In particular it has been shown that the morphological properties of a Gaussian random field are analytically known with non-Gaussian corrections giving rise to changes in the morphological properties of a random field. The MFs are therefore sensitive to non-Gaussianity allowing us to construct a statistical estimator based on analytically known results for a Gaussian random field. Using the formalism presented by \cite{Matsubara02} and \cite{Hikage06} it has been shown that the MFs, to leading order, depend on a set of skewness parameters $S_0, S_1, S_2$ that encapsulate the non-Gaussian corrections to the random field. In particular, for a weakly non-Gaussian random field, the skewness parameters, a set of one-point estimators, reduce to a weighted probe of the bispectrum. The currently adopted approach has been to collapse all the information into the one-point estimators but at a loss of the ability to effectively discriminate between different types of non-Gaussianity. The shape of the bispectrum has been strongly emphasized as an extremely important tool in discriminating between various models for an inflationary scenario as well as secondary non-Gaussianity such as gravitationally induced non-linearity. Motivated by this we have presented a generalisation of these one-point estimators to a set of skew-spectra $S^0_l, S^1_l, S^2_l$ that avoid collapsing all the information into a single estimator. It has been argued that by investigating the shape of the skew-spectra and it's relation to the bispectrum $B_{l_1 l_2 l_3}$ we can discriminate more effectively between the various bispectral configurations and ideally separate out the primordial and secondary contributions. 

These generalised skew-spectra are two-point statistics spatially but the dependence on the bispectrum is related to the fact that they are third (leading terms) order statistics in terms of non-Gaussianity. A cumulant correlator of order $p+q$ is constructed from the cross-correlation of the spatial fields (or the derivative fields) $\langle \Psi^p ({\bf{x}}) \Psi^q ({\bf{x}}) \rangle$. This cumulant correlator will probe the polyspectra of order $p+q$ \citep{Munshi09}. 

This paper has only considered the generalised skewness parameters up to $\mathcal{O} (\sigma_0)$ with the resulting spectra corresponding to a weighted probe of bispectrum. The formalism is developed in terms of a perturbative expansion about a Gaussian random field and we necessarily truncate the results to the desired order. We could consider higher order corrections with the results of $\mathcal{O} (\sigma^2_0)$ for which the resulting spectra would provide a weighted probe of the trispectrum (see e.g. \cite{Matsubara10,MunshivW11}). The next-to-leading order results are appropriately named kurt-spectra due to their relation to the more conventionally used kurtosis parameters. Although we have not considered the next-to-leading order non-Gaussian corrections, the unprecedented level of detail in upcoming surveys makes investigations of next-to-leading order statistics a promising future direction. 

The skew-spectra have been constructed from various products of the spatial and derivative fields (e.g. $\nabla \Psi, \nabla^2 \Psi, \nabla \Psi \cdot \nabla \Psi$, etc). The skew-spectra can be related to the spatially defined cumulant correlators and the skewness parameters can be constructed from the corresponding skew-spectra. Given the momenta-dependence of the skew-spectra it was shown that they carry greater discriminating power against the various bispectra configurations which can be compared to observational or numerical results. Something that has not yet been considered is the role of noisy data, systematics or survey masks. A discussion of error on the estimators and their scatter will be presented elsewhere. The presence of such noise will lead to scatter in our estimators necessitating a greater understanding of the systematics in upcoming LSS surveys (see e.g. \cite{Hikage11}). 

The systematic study of the skew-spectra presented in this paper focused on three different cosmological data sets: 3D density, 3D velocity and 2D projected. For each of these models we numerically investigated the shape of the bispectrum for three primordial models (local, equilateral and folded) along with the analytical gravitationally induced bispectrum. The primordial bispectra were shown to be dependent on the underlying matter power spectrum, transfer function with an overall amplitude set by the background cosmology and smoothing scale. The 2D skew-spectra were calculated for the innermost kernel in the integral with the integral over the background geometry neglected. The shape and amplitude of the underlying spectra is still of interest however.  

The approach to probing non-Gaussianity presented in this paper is just one of many areas of similar work. It is accepted that the CMB should provide one of the cleanest probes of the early Universe with the density perturbations being adequately modeled by linear perturbation theory. The non-linear nature of gravitational instability means that the primary source of non-Gaussianity in LSS or weak-lensing surveys is most likely non-primordial. Nevertheless, LSS surveys are currently thought to be able to place comparable constraints on primordial non-Gaussianity to that of the CMB. Likewise, recent developments on the analytical modeling of weak lensing has lead to the hope that upcoming weak lensing surveys could provide an unbiased probe of the underlying matter distribution yielding a cleaner probe of the statistics of gravitational clustering than conventional LSS surveys (e.g. \cite{MunshivW11}). Another possibility that could offer the cleanest probe of primordial non-Gaussianity is that of a non-Gaussian stochastic background of gravitational waves. The prospects for direct detection are not currently optimistic due to the weakness of the 3-point term in the graviton interactions \citep{Adshead09a} though it is hoped that indirect detection through, for example, CMB polarisation could be a possibility. 

An assumption that was made throughout this paper was the use of a Gaussian window function. This is a popular, though arbitrary, choice and we could equally have performed the analysis using alternative window functions. Another popular window function used in the literature is the top-hat window function \citep{Bernardeau02}. The formalism presented allows the results to be generalised to arbitrary functions but this is not thought to change the overall conclusions of the paper. 

In all models the gravitational instability bispectrum was an order of magnitude or two greater than that of the primordial bispectra. At very small wavenumbers the local primordial bispectra was starting to become comparable to the gravitational instability skewness. The results quoted are for $f_{\rm{NL}} \sim 20$ which may be somewhat optimistic with the true amplitude for non-Gaussianity lying closer to unity and as a result this would further reduce the skew-spectra amplitudes by an order of magnitude. Interestingly it seems that the shapes of the primordial bispectra are distinguishable even if the gravitational instability non-Gaussianity is the dominant contribution for all surveys discussed. The results do not take into account systematics or noise so no constraints on the signal-to-noise ratio may be discussed but a more detailed account will be presented elsewhere. The shape of the gravitational instability skew-spectra could be useful in quantifying predictions from perturbative treatments of gravitational instability, N-body simulations or semi-analytic approximations and it is worthwhile considering the prospects for disentangling any primordial contributions from gravitational instability.

We have studied the skew-spectra for the 3D galaxy distribution as well as 2D projected surveys. For the 3D surveys we ignored the effects of redshift space distortion. Redshift space effects are important as these correspond more directly to the observables in conventional large scale structure surveys. Examples of the role of redshift space distortion can be found in \citep{Kaiser87,Hamilton97,Shaw08,Okumara10,Sato11}. A more general treatment taking into account redshift space effects will be presented elsewhere. 

\section{Acknowledgements}
\label{acknow}
GP acknowledges support from an STFC studentship. 
DM acknowledges support
from STFC standard grant ST/G002231/1 at School of Physics and
Astronomy at Cardiff University where this work was completed. We thank Peter Coles and Ian Harrison for 
useful discussions. DM also wants to thank Alan Heavens, Joseph Smidt, Ludo van-Waerbeke and Patrick Valageas for many
useful discussions. We also thank an anonymous referee for useful comments and feedback on this paper. 
\appendix

\section{Special Functions and Their Properties}
\label{AppendixA}
A number of calculations in both this paper and key references make use of a number of properties 
of the special functions (e.g. the Legendre polynomials). In this section we provide some key results used in the derivations.

In order to perform the angular integration in the skewness parameters it is convenient 
to use the relationship between the Legendre polynomials and the modified Bessel 
functions. The origin of this relationship can be seen by performing an expansion of the angularly dependent Gaussian window function $W(k_3 R)$ \citep{Lokas95}:
\begin{equation}
 e^{- {\bf{l_1}} \cdot {\bf{l_2}}} = e^{- l_1 l_2 \mu} = \displaystyle\sum^{\infty}_{m = 0} \left( -1 \right)^m (2m + 1) 
 \sqrt{\frac{\pi}{2 l_1 l_2}} I_{m+1/2} (l_1 l_2) 
P_m (\mu) .
\end{equation}
The Legendre polynomials $P_m (\mu)$ inherit the angular dependence and upon integration we can use the orthogonality and completeness of these functions to 
simplify our results considerably:
\begin{equation}
 \int_{-1}^{1} d \mu P_m (\mu) P_n (\mu) = \frac{2 \delta_{m n}}{(2m + 1)} .
\end{equation}
This relationship motivates us to associate the angular terms dependent on $\mu$ to the appropriate Legendre polynomial. After a small amount of algebra we 
recover the formula provided in \cite{Matsubara02} relating the Legendre polynomials and modified Bessel functions:
\begin{equation}
 \int_{-1}^{1} d \mu P_{m} (\mu) e^{- l_1 l_2 \mu} = (-1)^m \sqrt{\frac{2 \pi}{l_1 l_2}} I_{m+1/2} (l_1 l_2) .
\end{equation}

\n
A useful set of relationships are those for the 2-dimensional case in which the recursion relationships allow us to perform the angular integration. The following formulas are particularly useful (see e.g. \cite{Matsubara02}):

\begin{align}
 &\frac{1}{\pi} \int \frac{d \mu}{\sqrt{1 - \mu^2}}       e^{- l_1 l_2 \mu} = I_0 (l_1 l_2) , \\
 &\frac{1}{\pi} \int \frac{d \mu}{\sqrt{1 - \mu^2}} \mu   e^{- l_1 l_2 \mu} = - I_1 (l_1 l_2) , \\
 &\frac{1}{\pi} \int \frac{d \mu}{\sqrt{1 - \mu^2}} \mu^2 e^{- l_1 l_2 \mu} = \frac{1}{2} \left( I_0 (l_1 l_2) + I_2 (l_1 l_2) \right) , \\
 &\frac{1}{\pi} \int \frac{d \mu}{\sqrt{1 - \mu^2}} \mu^3 e^{- l_1 l_2 \mu} = -\frac{1}{4} \left( 3 I_1 (l_1 l_2) + I_3 (l_1 l_2) \right), \\
 &\frac{1}{\pi} \int \frac{d \mu}{\sqrt{1 - \mu^2}} \mu^4 e^{- l_1 l_2 \mu} = \frac{1}{8} \left( 3 I_0 (l_1 l_2) + 4 I_2 (l_1 l_2) + I_4 (l_1 l_2) \right) .
\end{align}

%
These results were used in the text of the paper.
\section{Transfer Function}
\label{AppendixB}
The transfer function allows us to evolve an initial sub-Hubble mode to some final state at an arbitrary time. 

\begin{equation}
 T(k,a) = \frac{ D (a_i)}{\delta (k,a_i)} \frac{\delta(k,a)}{D (a)}
\end{equation}

\n
The growth function defines the linear growth of long wavelength perturbations \citep{Bardeen86}. 
We require that the initial time is such that it is before any scale of interest has become sub-Hubble. By definition we have:

\begin{equation}
 \Phi (k,a) = T(k,a) \phi(k,a_i)
\end{equation}

\n
This allows us to evolve the initially sub-Hubble modes from a given model for the early-Universe to make predictions for the power spectrum 
at an arbitrary time given that the approximation for the transfer function holds. The model we adopt for the transfer function is that given 
in \citep{Bardeen86} for a cold dark matter model with adiabatic initial conditions:

\begin{equation}
 T(q) = \frac{\ln{1 + 2.34q}}{2.34q} \left[ 1 + 3.89q + (16.1q)^2 + (5.46q)^3 + (6.71q)^4 \right]^{-1/4}
\end{equation}

\n
where we have adopted the following variables:

\begin{equation}
 q = \frac{k}{(\Gamma \rm{Mpc}^{-1})} ; \; \Gamma = \Omega_{m,0} h^2
\end{equation}

\n
This fitting function built on the work of \citep{Bardeen85,Bond83,Bond84,Efstathiou85}.

\bibliography{paper.bbl}

\begin{thebibliography}{99}
\bibitem[\protect\citeauthoryear{Adshead and Lim}{2009}]{Adshead09a} Adshead P., Lim E., 2010, PRD, D82, 024023
\bibitem[\protect\citeauthoryear{Adler}{1981}]{Adler81} Adler R. J., 1981, The Geometry of Random Fields, Chichester: Wiley
\bibitem[\protect\citeauthoryear{Dark Energy Task Force}{2006}]{Albrecht06} Albrecht A., Bernstein G., Cahn R., Freedman W. L., Hewitt J., Hu W., Huth J., Kamionkowski M., Kolb E. W., Knox L., Mather J. C., Staggs S., Suntzeff N. B., 2006, Report of the Dark Energy Task Force, arXiv:astro-ph/0609591v1
\bibitem[\protect\citeauthoryear{Allen et al.}{1987}]{Allen87} Allen T., Grinstein B., Wise M., 1987, Phys. Lett. B, 197, 66
\bibitem[\protect\citeauthoryear{Babich}{2005}]{Babich05} Babich D., 2005, PRD, D72, 043003
\bibitem[\protect\citeauthoryear{Bardeen}{1985}]{Bardeen85} Bardeen J. M., 1985, Proc. Inner Space/Outer Space Conf., ed. Kolb E. W., Turner M. S., 
Olive K., Seckel D. and Lindley D., Chicago: University of Chicago Press)
\bibitem[\protect\citeauthoryear{Bardeen et al.}{1986}]{Bardeen86} Bardeen, J. M., Bond, J. R., Kaiser, N., \& Szalay, A. S. 1986, ApJ, 304, 15
\bibitem[\protect\citeauthoryear{Baumann \& McAllister}{2009}]{Baumann09} Baumann D., McAllister L., 2009, Ann. Rev. Nucl. Part. Sci., 59
\bibitem[\protect\citeauthoryear{Baumann et al.}{2009}]{Baumann09b} Baumann D., et al., 2009, AIP Conf.Proc.1141:10-120
\bibitem[\protect\citeauthoryear{Bernardeua}{1994}]{Bernardeau94} Bernardeau F., 1994, ApJ., 433
\bibitem[\protect\citeauthoryear{Bernardeau}{2002}]{Bernardeau02} Bernardeau F., Colombi S., Gaztanaga E., Scoccimarro R., 2002, Phys. Rept., 367, 1
\bibitem[\protect\citeauthoryear{Blake et al}{2011}]{Blake11} Blake C., et al., 2011, MNRAS, 415, 3, 2892-2909
\bibitem[\protect\citeauthoryear{Bond and Efstathiou}{1984}]{Bond84} Bond J. R., Efstathiou G., 1984, ApJ Lett., 285, L45
\bibitem[\protect\citeauthoryear{Bond and Szalay}{1983}]{Bond83} Bond J. R., Szalay A. S., 1983, ApJ, 277, 443
\bibitem[\protect\citeauthoryear{Cabella et al. }{2006}]{Cabella06}
  Cabella P., Hansen F. K., Liguori M., Marinucci D., Matarrese S., Moscardini L., Vittorio N., 2006, MNRAS, 369, 819
\bibitem[\protect\citeauthoryear{Canavezes et al.}{1998}]{Canavezes98} Canavezes A., et al., 1998, MNRAS, 297, 777
\bibitem[\protect\citeauthoryear{Carbone, Verde and Matarrese}{2008}]{Carbone08} Carbone C., Verde L., Matarrese S., 2008, ApJ, 684:L1-L4
\bibitem[\protect\citeauthoryear{Chen}{2010}]{Chen10} Chen X., 2010, Adv. Astron., 2010:638979
\bibitem[\protect\citeauthoryear{Creminelli}{2003}]{Creminelli03}Creminelli P., 2003, JCAP 0310, 003
\bibitem[\protect\citeauthoryear{Creminelli et al.}{2006}]{Creminelli06} Creminelli P., Nicolis A., Senatore L., Tegmark M., Zaldarriaga M., 2006, JCAP, 0605, 4 
\bibitem[\protect\citeauthoryear{Creminelli et al.}{2007}]{Creminelli07a} Creminelli P., Senatore L., Zaldarriaga M., Tegmark M., 2007, JCAP, 3, 5 
\bibitem[\protect\citeauthoryear{Creminelli, Senatore, \& Zaldarriaga}{2007}]{Creminelli07b} Creminelli P., Senatore L., Zaldarriaga M., 2007, JCAP, 3, 19 
\bibitem[\protect\citeauthoryear{Coles}{1988}]{Coles88} Coles P., 1988, MNRAS, 234, 509
\bibitem[\protect\citeauthoryear{Coles \& Plionis}{1991}]{Coles91} Coles P., Plionis M., 1991, MNRAS, 250, 75. 
\bibitem[\protect\citeauthoryear{Colley, Gott and Park}{1996}]{Colley96} Colley W. N., Gott J. R., Park C., 1996, MNRAS, 281, 4, L82-L84
\bibitem[\protect\citeauthoryear{The COrE Collaboration}{2011}]{Core11} The COrE Collaboration, 2011, arXiv:1102.2181v2
\bibitem[\protect\citeauthoryear{Cooray}{2001}]{Cooray01} Cooray A., PRD, D79
\bibitem[\protect\citeauthoryear{Dalal et al.}{2008}]{Dalal08} Dalal N., et al., 2008, PRD, D77, 123514
\bibitem[\protect\citeauthoryear{Danielsson}{2002}]{Danielsson02} Danielsson U. H., 2002, PRD, D66, 023511
\bibitem[\protect\citeauthoryear{de Troia et al.}{2007}]{deTroia07} de Troia G., et al., 2007, ApJ, 670, L36
\bibitem[\protect\citeauthoryear{Easther et al.}{2001}]{Easther01} Easther R., Greene B. R., Kinney W. H., Shiu G., 2001, PRD, D64, 103502
\bibitem[\protect\citeauthoryear{Efstathiou and Bond}{1985}]{Efstathiou85} Efstathiou G., Bond J. R., 1985, MNRAS, 
\bibitem[\protect\citeauthoryear{Eisenstein and Hu}{1998}]{Eisenstein97a} Eisenstein D. J., Hu W., 1998, ApJ, 496, 605
\bibitem[\protect\citeauthoryear{Eisenstein and Hu}{1999}]{Eisenstein97b} Eisenstein D. J., Hu W., 1999, ApJ, 511, 5
\bibitem[\protect\citeauthoryear{Eriksen et al.}{2004}]{Eriksen04} Eriksen H. K., et al., 2004, ApJ, 612, 64
\bibitem[\protect\citeauthoryear{Fry and Gaztanaga}{1993}]{Fry93} Fry J. N., Gaztanaga E., 1993, ApJ, 413, 447-452
\bibitem[\protect\citeauthoryear{Gangui}{1994}]{Gangui94} Gangui A., Lucchin F., Matarrese S., Mollerach S., 1994, ApJ, 430, 447
\bibitem[\protect\citeauthoryear{Gleser et al.}{2006}]{Gleser06} Gleser L., Nusser A., Ciardi B., Desjacques V., 2006, MNRAS, 370, 1329-1338
\bibitem[\protect\citeauthoryear{Gott et al.}{1986}]{Gott86} Gott J. R., Mellot A. L., Dickinson M., 1986, ApJ, 306, 341
\bibitem[\protect\citeauthoryear{Gott et al.}{1989}]{Gott89} Gott J. R., et al., 1989, ApJ, 340, 625
\bibitem[\protect\citeauthoryear{Gott et al.}{1990}]{Gott90} Gott J. R., et al., 1990, ApJ, 352, 1-14
\bibitem[\protect\citeauthoryear{Gott et al.}{1992}]{Gott92} Gott J. R., Mao S., Park C., Lahav O., 1992, ApJ, 385, 26
\bibitem[\protect\citeauthoryear{Green et al.}{2011}]{Green11} Green J., et al., 2011, arXiv:1108.1374v1
\bibitem[\protect\citeauthoryear{Guth}{1982}]{Guth82} Guth A., Pi S.-Y., 1982, Phys. Rev. Lett., 49, 1110
\bibitem[\protect\citeauthoryear{Hadwiger}{1959}]{Hadwiger59} Hadwiger H., 1959, Normale Koper im Euclidschen raum und ihre topologischen and metrischen 
    Eigenschaften, Math Z., 71, 124
\bibitem[\protect\citeauthoryear{Hamilton}{1992}]{Hamilton92} Hamilton A. J. S., 1992, ApJ, 385, L5-L8
\bibitem[\protect\citeauthoryear{Hamilton}{1997}]{Hamilton97} Hamilton A. J. S., 1998, The Evolving Universe, edited by D. Hamilton, Vol. 231 of Astro. and Space Science Library, 185-275, arXiv:astro-ph/9708102v2
\bibitem[\protect\citeauthoryear{Hawking}{1982}]{Hawking82} Hawking S.W., 1982, Phys. Lett. B, 115, 295
\bibitem[\protect\citeauthoryear{Heavens}{1998}]{Heavens98} Heavens A., 1998, MNRAS, 299, 3
\bibitem[\protect\citeauthoryear{Hikage et al.}{2008}]{Hikage08} Hikage C., et al., 2008, MNRAS, 385, 1613-1620
\bibitem[\protect\citeauthoryear{Hikage, Komatsu \& Mastubara}{2006}]{Hikage06} Hikage C., Komatsu E., Matsubara T., 2006, ApJ, 653, 11
\bibitem[\protect\citeauthoryear{Hikage et al.}{2008}]{HikageM08} Hikage C., et al., 2008, MNRAS, 389, 1439
\bibitem[\protect\citeauthoryear{Hikage, Taruya and Suto}{2003}]{HikageT03} Hikage C., Taruya A., Suto Y., 2003, Publ. Astron. Soc. Jap., 55, 335
\bibitem[\protect\citeauthoryear{Hikage et al.}{2002}]{Hikage02} Hikage C., et al., 2002, Publ. Astron. Soc. Jap., 54, 707
\bibitem[\protect\citeauthoryear{Hikage et al.}{2003}]{Hikage03} Hikage C., et al., 2003, Publ. Astron. Soc. Jap., 55, 911
\bibitem[\protect\citeauthoryear{Hikage et al.}{2011}]{Hikage11} Hikage C., Takada M., Spergel D. N., 2011, arXiv.1106.1640v1
\bibitem[\protect\citeauthoryear{Jeong and Komatsu}{2009a}]{Jeong09a} Jeong D., Komatsu E., 2009a, ApJ, 691, 569
\bibitem[\protect\citeauthoryear{Jeong and Komatsu}{2009b}]{Jeong09b} Jeong D., Komatsu E., 2009b, ApJ, 703, 1230
\bibitem[\protect\citeauthoryear{Kachru et al.}{2003}]{Kachru03} Kachru S., et al., 2003, JCAP, 10, 013
\bibitem[\protect\citeauthoryear{Kaiser}{1987}]{Kaiser87} Kaiser N., 1987, MNRAS, 227, 1-21
\bibitem[\protect\citeauthoryear{Kerscher et al.}{2001}]{Kerscher01} Kerscher M., et al., 2001, A \& A, 373, 1-11
\bibitem[\protect\citeauthoryear{Komatsu et al.}{2003}]{Komatsu03} Komatsu E., et al., 2003, ApJS, 148, 119
\bibitem[\protect\citeauthoryear{Komatsu et al.}{2011}]{Komatsu11} Komatsu E., et al., 2011, ApJS, 192
\bibitem[\protect\citeauthoryear{Komatsu, Spergel \& Wandelt}{2005}]{Komatsu05} Komatsu E., Spergel D. N., Wandelt B. D., 2005, ApJ, 634, 14 
\bibitem[\protect\citeauthoryear{Komatsu et al.}{2001}]{Komatsu01} Komatsu E., Spergel D., 2001, PRD, D63, 063002
\bibitem[\protect\citeauthoryear{Laureijs et al.}{2009}]{Laureijs09} Laureijs R., et al., Euclid Science Study Team, 2009, ESA/SRE(2009)2, arXiv:0912.0914v1
\bibitem[\protect\citeauthoryear{Linde}{2005}]{Linde05} Linde A., 2005, Contemp. Concepts Phys., 5
\bibitem[\protect\citeauthoryear{Lokas et al.}{1995}]{Lokas95} Lokas E. L., Juszkiewicz R., Weinberg D. H., Bouchet F. R., 1995, MNRAS, 274, 3
\bibitem[\protect\citeauthoryear{Lyth \& Riotto}{1999}]{Lyth99} Lyth D. H., Riotto A., 1999, Phys. Rept., 314
\bibitem[\protect\citeauthoryear{Ma et al.}{1999}]{Ma99} Ma C-P., Caldwell, R. R., Bode P., Wang L., 1999, Astrophys. J. Lett., 521L, 1
\bibitem[\protect\citeauthoryear{Maldacena}{2003}]{Maldacena03} Maldacena J. 2003, JHEP, 0305, 013
\bibitem[\protect\citeauthoryear{Martin \& Brandenberger}{2001}]{Martin01} Martin J., Brandenberger R. H., 2001, PRD, D63, 123501
\bibitem[\protect\citeauthoryear{Matarrese and Verde}{2008}]{Matarrese08} Matarrese S., Verde L., 2008, ApJ, 677:L77-L80
\bibitem[\protect\citeauthoryear{Matsubara}{1994}]{Matsubara94} Matsubara T., 1994, Astrophys. J. Lett., 434, L43
\bibitem[\protect\citeauthoryear{Matsubara}{1995}]{Matsubara95} Matsubara T., 1995, ApJS, 101, 1
\bibitem[\protect\citeauthoryear{Matsubara}{2000}]{Matsubara00} Matsubara T., 2000, ApJ, 535, 1-23
\bibitem[\protect\citeauthoryear{Matsubara and Jain}{2001}]{Matsubara01} Matsubara T., Jain B., 2001, ApJ, 552, L89
\bibitem[\protect\citeauthoryear{Matsubara}{2002}]{Matsubara02} Matsubara T., 2003, ApJ, 584, 1
\bibitem[\protect\citeauthoryear{Matsubara}{2010}]{Matsubara10} Matsubara T., 2010, PRD, D81, 083505
\bibitem[\protect\citeauthoryear{Matsubara \& Yokohama}{1996}]{MatsubaraY96} Matsubara T., Yokohama J., 1996, ApJ, 463
\bibitem[\protect\citeauthoryear{Mazumdar \& Rocher}{2011}]{Mazumdar11} Mazumdar A. Rocher J., 2011, Phys. Rept., 497, 85
\bibitem[\protect\citeauthoryear{McDonald}{2006}]{McDonald06} McDonald P., 2006, PRD, D10, 103512
\bibitem[\protect\citeauthoryear{McDonald}{2008}]{McDonald08} McDonald P., 2008, PRD, D12, 123519
\bibitem[\protect\citeauthoryear{Mead, Lewis and King}{2011}]{Mead11} Mead J. M. G., Lewis A., King L., 2011, PRD, D83, 023507
\bibitem[\protect\citeauthoryear{Mecke et al.}{1994}]{Mecke94} Mecke K. R., Buchert T., Wagner H., 1994, A \& A, 288, 697
\bibitem[\protect\citeauthoryear{Medeiros \& Contaldi}{2006}]{Medeiros06}Medeiros J., Contaldi C.R, 2006, MNRAS, 367, 39
\bibitem[\protect\citeauthoryear{Melott}{1990}]{Melott89} Melott A. L., 1990, Phys. Rep., 193, 1
\bibitem[\protect\citeauthoryear{Moore et al.}{1992}]{Moore92} Moore B., et al., 1992, MNRAS, 256, 477
\bibitem[\protect\citeauthoryear{Mukhanov and Chibisov}{1981}]{Mukhanov81} Mukhanov V.F., Chibisov, G.V., 1981, Sov. Phys. - JETP 33, 532
\bibitem[\protect\citeauthoryear{Munshi \& Heavens}{2009}]{Munshi09} Munshi D., Heavens A., 2010, MNRAS, 401, 2406
\bibitem[\protect\citeauthoryear{Munshi, Smidt and Cooray}{2010}]{Munshi10} Munshi D., Smidt J., Cooray A., 2010, arXiv:1011.5224v1
\bibitem[\protect\citeauthoryear{Munshi et al.}{2011}]{Munshi11} Munshi D., Coles P., Cooray A., Heavens A., Smidt J., 2011, MNRAS, 410, 1295
\bibitem[\protect\citeauthoryear{Munshi et al.}{2011}]{MunshivW11} Munshi D., van Waerbeke L., Smidt J., Coles P., 2011, arXiv:1103.1876v1
\bibitem[\protect\citeauthoryear{Munshi et al.}{2011}]{MunshiSZ11} Munshi D., Smidt J., Joudaki S., Coles P., 2011, arXiv:1105.5139v1
\bibitem[\protect\citeauthoryear{Natoli et al.}{2010}]{Natoli10} Natoli P., et al., 2010, 408, 3, 1658-1665
\bibitem[\protect\citeauthoryear{Nicholson and Contaldi}{2009}]{Nicholson09} Nicholson G., Contaldi C. R., 2009, JCAP, 0907:011 
\bibitem[\protect\citeauthoryear{Novikov, Schmalzing and Mukhanov}{2000}]{Novikov00} Novikov D., Schmalzing J., Mukhanov V. F., 2000, A \& A, 364
\bibitem[\protect\citeauthoryear{Okumara and Jing}{2011}]{Okumara10} Okumara T., Jing Y. P., 2011, ApJ, 726, 5
\bibitem[\protect\citeauthoryear{Park \& Gott}{1988}]{Park88} Park C., Gott J. R., 1988, BAAS, 20, 987
\bibitem[\protect\citeauthoryear{Park et al.}{2005}]{Park05} Park C., et al., 2005, ApJ, 633, 11
\bibitem[\protect\citeauthoryear{Peebles}{1980}]{Peebles80} Peebles P. J. E., 1980, ''The Large Scale Structure of the Universe'', Princeton University Press
\bibitem[\protect\citeauthoryear{Percival and White}{2009}]{Percival09} Percival W. J., White M., 2009, MNRAS, 393, 297
\bibitem[\protect\citeauthoryear{Percival et al.}{2007}]{Percival07} Percival W. J., et al., 2007, ApJ, 657, 51
\bibitem[\protect\citeauthoryear{The Planck Collaboration}{2006}]{PC06}The Planck Collaboration, 2006, ESA-SCI(2005)1, astro-ph/0604069
\bibitem[\protect\citeauthoryear{Polenta et al.}{2002}]{Polenta02} Polenta G., et al., 2002, ApJ, 572, L27-L32
\bibitem[\protect\citeauthoryear{Rhoads, Gott \& Postman}{1994}]{Rhoads94} Rhoads J. E., Gott J.R., Postman M., 1994, ApJ, 421, 1
\bibitem[\protect\citeauthoryear{Sahni, Sathyaprakash and Shandarin}{1998}]{Sahni98} Sahni V., Sathyaprakash B. S., Shandarin S. F., 1998, ApJ, 495, L5-L8
\bibitem[\protect\citeauthoryear{Salopek and Bond}{1990}]{Salopek90} Salopek D. S., Bond J. R., 1990, PRD, D42, 3936
\bibitem[\protect\citeauthoryear{Sato et al.}{2001}]{Sato01} Sato J., et al., 2001, ApJ, 421, 1
\bibitem[\protect\citeauthoryear{Sato and Matsubara}{2011}]{Sato11} Sato J., Matsubara T., 2011, PRD, D84, 043501
\bibitem[\protect\citeauthoryear{Schlegel et al}{2010}]{Schlegel10} Schlegel D., et al., 2010, The BigBOSS Experiment, arXiv:1106.1706v1
\bibitem[\protect\citeauthoryear{Schmalzing \& Buchert}{1997}]{Schmalzing97} Schmalzing J., Buchert T., 1997, ApJ, 482, L1
\bibitem[\protect\citeauthoryear{Schmalzing \& Diaferio}{2000}]{Schmalzing00} Schmalzing J., Diaferio A., 2000, MNRAS, 312
\bibitem[\protect\citeauthoryear{Schmalzing \& G\'{o}rski}{1998}]{Schmalzing98} Schmalzing J., G\'{o}rski K. M., 1998, MNRAS, 297, 355
\bibitem[\protect\citeauthoryear{Schmalzing et al.}{1999}]{Schmalzing99} Schmalzing J., et al., 1999, ApJ., 526, 568-578
\bibitem[\protect\citeauthoryear{Scoccimaro}{2000}]{Scoccimarro00} Scoccimarro R., 2000, ApJ, 544, 597-615
\bibitem[\protect\citeauthoryear{Scoccimarro et al.}{2004}]{Scoccimarro04} Scoccimaro R., Sefusatti E., Zaldarriaga M., 2004, PRD, D69, 103513
\bibitem[\protect\citeauthoryear{Seery and Lidsey}{2005}]{Seery05} Seery D., Lidsey J. E., JCAP, 0605, 45
\bibitem[\protect\citeauthoryear{Sefusatti and Komatsu}{2007}]{Sefusatti07} Sefusatti E., Komatsu E., 2007, PRD, D76, 083004
\bibitem[\protect\citeauthoryear{Shaw and Lewis}{2008}]{Shaw08} Shaw J. R., Lewis A., 2008, PRD, D78, 103512
\bibitem[\protect\citeauthoryear{Smith, Senatore \& Zaldarriaga}{2009}]{Smith09} Smith K. M., Senatore L., Zaldarriaga M., 2009, JCAP, 0909:006
\bibitem[\protect\citeauthoryear{Starobinksy}{1982}]{Starobinsky82} Starobinksy A.A., 1982, Phys. Lett. B, 117, 175
\bibitem[\protect\citeauthoryear{Taruya et al.}{2002}]{Taruya02} Taruya A., et al., 2002, ApJ, 571, 638
\bibitem[\protect\citeauthoryear{Tegmark et al.}{2004}]{Tegmark04} Tegmark M., et al., 2004, PRD, D69, 103501
\bibitem[\protect\citeauthoryear{Tegmark et al.}{2004a}]{Tegmark04a} Tegmark M., et al., 2004, ApJ, 606, 702-740
\bibitem[\protect\citeauthoryear{Tomita}{1986}]{Tomita86} Tomita H., 1986, Progr. Theor. Phys., 76, 952
\bibitem[\protect\citeauthoryear{Verde et al.}{2000}]{Verde00} Verde L., Wang L., Heavens A. F., Kamionkowski M., 2000, MNRAS, 313, 141
\bibitem[\protect\citeauthoryear{Vogeley et al.}{1994}]{Vogeley94} Vogeley M. S., Park C., Geller M. J., Huchra J. P., Gott J. R., 1994, ApJ, 420, 525
\bibitem[\protect\citeauthoryear{Wang, Spergel and Strauss}{1999}]{Wang99} Wang Y., Spergel D. N., Strauss M. A., 1999, ApJ, 510, 20
\bibitem[\protect\citeauthoryear{Winitzki and Kosowksy}{1998}]{Winitzki98} Winitzki S., Kosowsky A., 1998, New. Astron., 3, 75
\end{thebibliography}
\end{document}